\documentclass[aps,prd,twocolumn,reprint,superscriptaddress,showkeys,showpacs,nofootinbib]{revtex4-2}
\usepackage[T1]{fontenc}
\usepackage[american]{babel}
\usepackage{epsfig}
\usepackage{xspace}
\usepackage{graphicx}
\usepackage{booktabs}
\usepackage{multirow}
\usepackage{dcolumn}
\usepackage{amsmath}
\usepackage{subfigure}
\usepackage{mathtools}
\usepackage{amsfonts}
\usepackage{amssymb}
\usepackage{epstopdf}
\usepackage{siunitx}
\usepackage{braket}
\usepackage{enumitem}
\usepackage{soul}
\usepackage{pifont}
\usepackage[normalem]{ulem}

\usepackage{hyperref}
\hypersetup{
    colorlinks=true, 
    linkcolor=blue, 
    citecolor=red
    }
    
\usepackage{xcolor}

\begin{document}
\title{Hyperonic compact stars with vector portal dark matter}

\author{Prafulla K.~Panda}
\email{prafulla.k.panda@gmail.com}
\affiliation{Department of Physics, Utkal University, Bhubaneswar-751004, India}

\author{Deepak Kumar}
\email{deepakk@iiserbpr.ac.in}
\affiliation{CFisUC, Department of Physics, University of Coimbra, PT 3004-516 Coimbra, Portugal}
\affiliation{Department of Physics, Indian Institute of Science Education and Research Berhampur, 760003, India}

\author{Hiranmaya Mishra}
\email{hiranmaya@niser.ac.in}
\affiliation{School of Physical Sciences, National Institute of Science Education and Research, An OCC of Homi Bhabha National Institute, Jatni - 752050, India}
\affiliation{Institute of Physics Bhubaneswar, Sachivalaya Marg, Bhubaneswar 751005, India}

\author{Sudhanwa Patra}
\email{sudhanwa@iitbhilai.ac.in}
\affiliation{Department of Physics, Indian Institute of Technology Bhilai, Durg 491002, India}

\begin{abstract} 
The appearance of hyperons in the core of neutron stars generally softens the equation of state (EOS), posing a longstanding challenge to the existence of observed two-solar-mass compact stars. We investigate whether repulsive interactions mediated by a dark-sector vector portal can provide an additional source of high-density pressure and thereby modify the structure of hyperonic compact stars. The baryonic sector is described within the modified quark-meson coupling (MQMC) model, in which the octet baryons are treated as confined relativistic constituent-quark systems interacting self-consistently through the $\sigma$, $\omega$, and $\rho$ fields within a mean field approximation. The dark sector consists of a fermionic dark matter coupled to baryonic matter through a neutral vector mediator $Z^\prime$, generating an additional repulsive contribution to the dense-matter EOS. We investigate the resulting equation of state, mass--radius relation, tidal deformability, and moment of inertia. The resulting mass--radius relations satisfy the observational bounds from massive pulsars, including  PSR J0740 + 6620, with maximum neutron star masses reaching approximately $2 M_{\odot}$. The resulting changes in tidal and rotational observables provide additional avenues for testing the dark-sector interaction through multimessenger observations.
\end{abstract}

\maketitle

\section{Introduction}
\label{sec:introduction}
The discovery of massive neutron stars with masses close to or exceeding $2\,M_{\odot}$ has provided stringent constraints on the equation of state (EOS) of dense matter. In particular, the precise mass measurements of the millisecond pulsars PSR J1614--2230 with a mass of $1.97\pm 0.04\,M_{\odot}$ obtained from Shapiro delay observations~\cite{Demorest2010}, and PSR J0348+0432 with mass of $M = 2.01 \pm 0.04\,M_{\odot}$ measured through combined optical and radio observations~\cite{Antoniadis2013}, have established the existence of very massive compact stars in nature. More recently, gravitational-wave observations, GW170817, from binary neutron star mergers have further constrained the mass--radius relation, tidal deformability, and high-density behavior of neutron star matter~\cite{LIGOScientific:2017vwq, LIGOScientific:2017ync}. These observations provide a unique opportunity to probe the properties of strongly interacting matter under extreme density conditions inaccessible in terrestrial experiments.


The internal structure of neutron stars is expected to be highly nontrivial, with densities in the central core reaching several times the nuclear saturation density~\cite{Haensel2003}. Under such conditions, the appearance of additional degrees of freedom beyond nucleons becomes energetically favorable. In particular, hyperons are expected to populate the inner core once the neutron chemical potential exceeds the hyperon rest masses. However, the inclusion of hyperons generally softens the EOS substantially, resulting in a reduction of the maximum stellar mass. This long-standing tension between the appearance of hyperons and the existence of two-solar-mass neutron stars is commonly referred to as the ``hyperon puzzle.'' 


A variety of theoretical approaches have been proposed to resolve this issue, including density-dependent hadronic interactions, nonlinear meson couplings, repulsive hyperon interactions, and phase transitions to deconfined quark matter at high density~\cite{Glendenning1985, GlendenningMoszkowski1991, Knorren1995, BalbergGal1997, Prakash1997, SchrammZschiesche2003, Long2012, Taurines2000, Lastowiecki2012}. Despite significant progress, the microscopic properties of dense hyperonic matter remain uncertain, particularly due to limited experimental information on hyperon--nucleon and hyperon--hyperon interactions.


In recent years, the possibility that dark matter may accumulate inside compact stars has attracted considerable attention. Compact stars provide a natural environment for probing dark matter interactions because of their extremely high density and strong gravitational fields. Depending on the nature of the dark sector interactions, dark matter can modify the EOS, alter the stellar composition, and significantly affect observable quantities such as the maximum mass, radius, tidal deformability, and moment of inertia of neutron stars. Among various dark matter scenarios, vector portal models have emerged as particularly interesting candidates, where fermionic dark matter interacts with standard model particles through a neutral vector mediator. Such models can generate additional repulsive interactions at high density and may help stiffen the EOS even in the presence of hyperons.


In the present work, we investigate compact star matter containing hyperons and vector portal dark matter within the framework of the modified quark-meson coupling (MQMC) model~\cite{Batista2002}. In this approach, baryons are described as systems of three independent relativistic constituent quarks confined by an effective phenomenological potential with an equally mixed scalar--vector harmonic oscillator structure. Unlike the conventional bag-model-based quark-meson coupling approach, the MQMC model incorporates quark confinement through a relativistic independent quark potential, allowing a more realistic treatment of the internal baryon structure. Earlier studies have shown that such confinement schemes lead to successful descriptions of symmetric and asymmetric nuclear matter properties, finite nuclei, and hyperonic matter \cite{BarikDash1986a, BarikDash1986b, Guichon1988, Guichon1996, Panda1997, Frederico1989, Whittenbury2014, Panda2004, Barik2013, Mishra2015}.


In the present formulation, the baryon--baryon interaction is generated self-consistently through the coupling of quarks to the scalar $\sigma$, vector $\omega$, and isovector $\rho$ mesons within the mean-field approximation. The effective baryon masses in the medium are obtained by incorporating the center-of-mass correction, pionic correction associated with chiral symmetry restoration, and one-gluon exchange contributions. In addition, we include a nonlinear $\omega$--$\rho$ interaction term in order to investigate its influence on the density dependence of the symmetry energy and neutron star properties.


Most previous investigations of dark matter admixed compact stars have focused either on two-fluid descriptions, where dark matter interacts only gravitationally with baryonic matter, or on single-fluid models involving scalar portal interactions between dark matter and nucleons. Scalar mediators primarily modify the effective nucleon mass through attractive interactions and generally soften the EOS. An equally compelling but comparatively less explored possibility is the vector portal, in which fermionic dark matter interacts with Standard Model quarks through a neutral vector mediator $Z^\prime$~\cite{Patra:2016ofq,Patra:2016shz,Taramati:2024kkn,Patel:2024zsu}. Unlike scalar interactions, vector mediators modify the chemical potentials of both baryons and dark matter, generating additional repulsive interactions at high density. Such repulsion naturally stiffens the EOS and therefore provides a promising mechanism for counterbalancing the softening caused by hyperons. Furthermore, the same $Z^\prime$ mediator responsible for dense matter interactions establishes a direct connection between compact-star phenomenology and terrestrial searches through direct detection, indirect detection, and collider experiments.

The hyperon couplings are constrained using empirical hypernuclear optical 
potentials at saturation density. In particular, we adopt the values
$U_\Lambda=-28~{\rm MeV}$, $U_\Sigma=30~{\rm MeV}$, $U_\Xi=-18~{\rm MeV}$
while also exploring different choices of the $\Xi$ potential because of its 
remaining experimental uncertainty. The dark matter sector is modeled through 
a fermionic dark matter particle interacting with baryonic matter via a vector 
mediator portal. We then investigate the combined effects of hyperons, nonlinear 
meson interactions, and dark matter on the EOS and the global properties of 
compact stars.

Recently, we investigated the effects of fermionic vector-portal dark matter on nucleonic neutron stars within the relativistic mean-field framework and demonstrated that neutron-star observations can place meaningful constraints on the dark matter and mediator parameters while remaining consistent with current multimessenger observations. Motivated by these findings, the present work extends the study to the more realistic and challenging case of hyperonic matter, where the competition between hyperon-induced softening and vector-mediated dark matter repulsion can play a crucial role in determining the structure of compact stars.


This paper is organized as follows. In Sec.~\ref{sec:MQMC}, we briefly 
outline the modified quark-meson coupling model and discuss the effective baryon 
mass corrections in dense matter. In Sec.~\ref{sec:eos}, we construct the EOS 
for hyperonic matter with vector portal dark matter under conditions of 
$\beta$ equilibrium and charge neutrality. Section~\ref{sec:ns_tidal} presents 
the formalism used to determine the neutron star structure, tidal deformability, 
and moment of inertia. The numerical results and discussions are presented in 
Sec.~\ref{sec:results}. Finally, we summarize our main conclusions in 
Sec.~\ref{sec:conclusion}.

\section{Modified Quark-Meson Coupling Model}
\label{sec:MQMC}
In conventional relativistic mean-field (RMF) models for nuclear matter where the baryons are generally treated as point-like particles, the modified quark--meson coupling (MQMC)~\cite{Batista2002} model incorporates the composite quark structure of baryons into the description of dense matter. In MQMC approach, a baryon is treated as a composite system of three relativistic constituent quarks confined by an effective potential, while the interaction of the quarks with the surrounding nuclear medium is generated self-consistently through the exchange of scalar and vector mesons. The MQMC framework has been applied to symmetric and asymmetric nuclear matter, finite nuclei, and compact stars, and provides a convenient microscopic description of hyperonic matter in which the internal structure of the baryons is retained explicitly ~\cite{Batista2002,BarikDash1986a, BarikDash1986b, Guichon1988, Guichon1996, Panda1997, Frederico1989, Whittenbury2014, Panda2004, Barik2013, Mishra2015}. 

In the present work, we consider the complete baryon octet, $B = \left\{
N,\Lambda,\Sigma,\Xi \right\}$,  embedded in uniform matter. The constituent quarks interact with the scalar $\sigma$, vector $\omega$, and isovector-vector $\rho$ mean fields. The scalar field modifies the effective quark mass and hence the internal structure of the baryons, whereas the vector fields generate the corresponding repulsive and isospin-dependent interactions. The nonlinear $\omega$--$\rho$ coupling is retained in the mean-field description in order to control the density dependence of the isovector sector. The resulting in-medium baryon masses are subsequently used to construct the equation of state of $\beta$-equilibrated hyperonic matter.

The confining interaction is chosen in an equally mixed scalar--vector harmonic oscillator form~\cite{Barik2013},
\begin{equation}
U(r)
=
\frac{1}{2}
\left(1+\gamma^0\right)
V(r),
\label{eq:confpot}
\end{equation}
where
\begin{equation}
V(r)=ar^2+V_0,
\qquad a>0.
\label{eq:potential}
\end{equation}
Here, $a$ and $V_0$ are the confinement parameters which determine the strength and depth of the potential, respectively. This confining interaction provides the zeroth-order quark dynamics inside the baryon. 
In the presence of the nuclear medium, the quark field $\psi_q(\vec r)$ (with  flavor index q) satisfies the Dirac equation
\begin{align}
\Big[
&
\gamma^0
\left(
\epsilon_q
-
V_\omega
-
\frac{1}{2}\tau_{3q}V_\rho
\right)
-
\vec{\gamma}\cdot\vec{p}
\nonumber\\
&
-
\left(
m_q-V_\sigma
\right)
-
U(r)
\Big]
\psi_q(\vec r)
=0,
\label{eq:dirac}
\end{align}
with $V_\sigma=g_\sigma^q\sigma_0$, $V_\omega=g_\omega^q\omega_0$, and $V_\rho=g_\rho^q b_{03}$. Here, $\sigma_0$, $\omega_0$, and $b_{03}$ denote the classical mean fields associated with the $\sigma$, $\omega$, and $\rho$ mesons, respectively. The quantities $g_\sigma^q$, $g_\omega^q$, and $g_\rho^q$ represent the corresponding quark--meson coupling constants. Furthermore, $m_q$ is the constituent quark mass and $\tau_{3q}$ denotes the third component of the isospin operator acting on the quark field.

For convenience, we define the effective quark energy and mass as
\begin{equation}
\epsilon_q^*
=
\epsilon_q
-
V_\omega
-
\frac{1}{2}\tau_{3q}V_\rho,
\qquad
m_q^*
=
m_q-V_\sigma .
\label{eq:effmass}
\end{equation}
The shifted quark energy and mass parameters are then introduced as
\begin{equation}
\epsilon_q'
=
\epsilon_q^*-\frac{V_0}{2},
\qquad
m_q'
=
m_q^*+\frac{V_0}{2}.
\label{eq:eprim}
\end{equation}
Following the standard MQMC formalism, we further define the quantities
\begin{equation}
\lambda_q
=
\epsilon_q'+m_q',
\qquad
r_{0q}
=
(a\lambda_q)^{-1/4},
\label{eq:lambda}
\end{equation}
where $r_{0q}$ corresponds to the effective oscillator length parameter for the confined quark. The ground-state quark energy is obtained from the eigenvalue condition
\begin{equation}
(\epsilon_q'-m_q')
\sqrt{\frac{\lambda_q}{a}}
=
3.
\label{eq:eigen}
\end{equation}

The solution of Eq.~(\ref{eq:eigen}) determines the effective quark energy $\epsilon_q^*$ in the medium. Consequently, the zeroth-order effective baryon mass is obtained by summing over the constituent quark energies, $E_B^{*0}=\sum_q\epsilon_q^*$. In addition to the zeroth-order quark energy, several important corrections must be incorporated in order to obtain the physical baryon mass in the medium. These include:
1) the spurious center-of-mass correction $\epsilon_{\rm c.m.}$,
2) the pionic correction $\delta M_B^\pi$ associated with chiral symmetry restoration, and 
3) the short-range one-gluon exchange corrections arising from color-electric and color-magnetic interactions. The detailed expressions for these corrections for the complete baryon octet are presented in Appendix~\ref{AppendixA}. Treating all corrections independently, the effective baryon mass in the medium is finally written as
\begin{equation}
M_B^*
=
E_B^{*0}
-
\epsilon_{\rm c.m.}
+
\delta M_B^\pi
+
(\Delta E_B)_g^{E}
+
(\Delta E_B)_g^{M}.
\label{eq:effective_mass}
\end{equation}
The medium-modified baryon masses obtained from Eq.~(\ref{eq:effective_mass}) are subsequently used to construct the equation of state for hyperonic compact star matter in the presence of dark matter interactions.
\section{Equation of State for Hyperonic Matter with Dark Matter}
\label{sec:eos}
We now construct the equation of state (EOS) of cold, charge-neutral,
$\beta$-equilibrated matter containing the complete baryon octet,
leptons, and fermionic dark matter. The baryonic sector is described
within the MQMC framework discussed in Sec.~\ref{sec:MQMC}, while the
interaction between the dark and baryonic sectors is mediated by a
neutral vector boson $Z^\prime_\mu$. The central feature of the vector
portal is that the temporal component of the mediator generates an
additional repulsive mean-field interaction. Consequently, the dark
sector can modify both the equilibrium composition and the stiffness
of the high-density EOS. The baryonic matter considered here consists of
\[ B=\{n,p,\Lambda,\Sigma^-,\Sigma^0,\Sigma^+,\Xi^-,\Xi^0\}, \]
together with electrons and muons. The dark sector is represented by a
fermionic particle $\chi$ with mass $m_\chi$. We work at zero
temperature, which is an appropriate approximation for cold neutron
stars whose interior temperature is much smaller than the relevant
Fermi energies. The effective Lagrangian density for the baryonic and leptonic sectors is written as
\begin{align}
\mathcal{L}_{\rm HM} ={}& \sum_B \overline{\psi}_B \Big[ i\gamma_\mu\partial^\mu
- M_B^*(\sigma) - g_{\omega B}\gamma_\mu\omega^\mu \nonumber \\
&- 
g_{\rho B}\gamma_\mu \pmb{I}_B\!\cdot\!\bm{b}^{\,\mu}\Big]\psi_B +
\frac{1}{2}\left(\partial_\mu\sigma\partial^\mu\sigma - m_\sigma^2\sigma^2 \right) \nonumber \\
&+ \frac{1}{2}m_\omega^2\omega_\mu\omega^\mu - \frac{1}{4}\Omega_{\mu\nu}\Omega^{\mu\nu} \nonumber \\
&+
\frac{1}{2}m_\rho^2 \pmb{b}_\mu\!\cdot\!\pmb{b}^{\,\mu} - \frac{1}{4} \pmb{B}_{\mu\nu}\!\cdot\!\pmb{B}^{\mu\nu} \nonumber \\
&+
\Lambda_v g_\omega^2g_\rho^2 \left(\omega_\mu\omega^\mu\right) \left(\pmb{b}_\nu\!\cdot\!\bm{b}^{\,\nu}\right) \nonumber \\
&+
\sum_{\ell=e,\mu} \bar{\psi}_{\ell} \left( i\gamma_\mu\partial^\mu - m_\ell \right)\psi_{\ell}\,,
\label{eq:hadronic_lagrangian}
\end{align}
where, $\Omega_{\mu\nu} = \partial_\mu\omega_\nu-\partial_\nu\omega_\mu$
and $\pmb{B}_{\mu\nu} = \partial_\mu\pmb{b}_\nu-\partial_\nu\pmb{b}_\mu$ are the field-strength tensors associated with the $\omega$ and
$\rho$ mesons, respectively. The quantity $M_B^*(\sigma)$ is the
density-dependent effective baryon mass obtained from the MQMC
calculation. Here, $\psi_B$ and $\psi_l$ denote the baryon and lepton fields, respectively, while $\vec I_B$ represents the isospin operator of baryon species $B$. The leptonic sector includes electrons and muons, which ensure charge neutrality and $\beta$ equilibrium inside compact stars. Within the mean-field approximation, the meson fields are replaced by their expectation values, namely $\sigma\rightarrow\sigma_0$, $\omega^\mu\rightarrow\delta^{\mu0}\omega_0$, and $\rho^\mu\rightarrow\delta^{\mu0}b_{03}$. 

The dark sector is introduced through a fermionic field $\chi$
nteracting with standard model quarks through a neutral vector mediator $Z^\prime_\mu$,
\begin{align}
\mathcal{L}_{\rm DM}
=&
\bar{\chi}
\left(
i\gamma_\mu\partial^\mu-m_\chi
\right)\chi
-
g_\chi
\bar{\chi}\gamma_\mu\chi Z^{\prime\mu}
\nonumber\\
&\hspace{-0.8cm} -
\sum_{q=u,d,s}
g_q\bar{q}\gamma_\mu q Z^{\prime\mu}
-\frac{1}{4}
Z^\prime_{\mu\nu}Z^{\prime\mu\nu}
+\frac{1}{2}
m_{Z^\prime}^2 Z^\prime_\mu Z^{\prime\mu},
\label{eq:dm_lagrangian}
\end{align}
where $g_\chi$ is the DM--$Z^\prime$ coupling, $g_q$ denotes the
quark--$Z^\prime$ coupling, and $m_\chi$ is the dark matter mass. 
At the hadronic level, the interaction between the vector mediator and
the baryons can be expressed in terms of an effective baryonic current,
\begin{equation}
\mathcal{L}_{Z^\prime B}
=
-
\sum_B
g_{BZ^\prime}
\bar{\psi}_B\gamma_\mu\psi_B
Z^{\prime\mu},
\label{eq:baryon_zprime_interaction}
\end{equation}
where $g_{BZ^\prime}$ denotes the effective baryon--$Z^\prime$
coupling obtained by matching the underlying quark-level interaction
onto the baryonic degrees of freedom. We keep these couplings general
in order to separate the EOS calculation from a particular ultraviolet
completion of the vector portal. 
A comment regarding the coupling of quarks to $Z'_\mu$ as in Eq.~(\ref{eq:dm_lagrangian})
may be relevant. This leads to affect the single quark energy $\epsilon_q^*$ in Eq.(\ref{eq:effmass}) as\begin{equation}
\epsilon_q^* = \epsilon_q - V_\omega - \frac{1}{2}\tau_{3q}V_\rho-g_qZ'_0.
\end{equation}
Such a modification of single quark energy due to portal mean field leads to a modification of effective mass of the baryons through Eq.(\ref{eq:eprim}) and Eq.(\ref{eq:effective_mass}). This  is in contrast to relativistic mean field models where vector portal mean field {\em does not} affect the effective  masses of the baryons.

For cold, homogeneous matter, only the temporal components of the
vector fields survive in the mean-field approximation. The equations of motion for the meson and vector mediator fields in uniform matter are obtained from the Euler--Lagrange equations and are given by
\begin{align}
m_\sigma^2\sigma_0 &= \sum_Bg_{\sigma B}C_B(\sigma)\rho_B^s,\label{eq:sigmafield}\\
{m_\omega^*}^2\omega_0 &= \sum_B g_{\omega B}\rho_B, \label{eq:omegafield}\\
{m_\rho^*}^2b_{03} &= \sum_B g_{\rho B}I_{3B}\rho_B, \label{eq:rhofield} \\
m_{Z'}^2Z_0' &= g_\chi\rho_\chi + \sum_B g_{BZ'}\rho_B. \label{eq:zfield}
\end{align}

The effective vector meson masses are modified by the nonlinear $\omega$--$\rho$ interaction term,
\begin{align}
{m_\omega^*}^2 &= m_\omega^2 + 2\Lambda_v g_\omega^2 g_\rho^2 b_{03}^2, \label{eq:mwstar} \\
{m_\rho^*}^2 &= m_\rho^2 + 2\Lambda_v g_\omega^2 g_\rho^2 \omega_0^2. \label{eq:mrhostar}
\end{align}
The quantity $g_{\sigma B}C_B(\sigma) = - \frac{\partial M_B^*(\sigma)}{\partial \sigma}$ encodes the scalar response of the baryon in the medium. The equation for the temporal component of the vector mediator is
\begin{equation}
m_{Z^\prime}^2 Z'_0 = g_\chi\rho_\chi + \sum_B g_{BZ^\prime}\rho_B.
\label{eq:zprime_field}
\end{equation}
Eq.~\eqref{eq:zprime_field} shows explicitly that the $Z^\prime$ mean field is sourced by both the dark-matter density and the baryon density. Consequently, even when the dark sector is dilute, its presence can influence the baryonic chemical potentials through the self-consistent vector field as seen in Eq.~\eqref{eq:chemical}.

At zero temperature, each fermionic species occupies states up to its corresponding Fermi momentum which define the number densities of specific species. The baryon number density is as follows
\begin{equation}
\rho_B = \frac{\gamma_B}{6\pi^2} k_{F,B}^3,
\label{eq:baryon_density}
\end{equation}
while the dark matter number density is follows, 
\begin{equation}
\rho_\chi = \frac{\gamma}{6\pi^2} k_{F\chi}^3.
\label{eq:dmdensity}
\end{equation}
where, $\gamma_B = \gamma =2$ are the spin degeneracy. The quantities $k_{F,B}$ and $k_{F\chi}$ represent the Fermi momenta of baryons and dark matter particles, respectively. The scalar density of baryon species $B$ is given by
\begin{equation}
\rho_B^s = \frac{\gamma}{2\pi^2} \int_0^{k_{F,B}} \frac{M_B^*}{\sqrt{k^2+M_B^{*2}}} k^2\,dk,
\label{eq:scalardensity}
\end{equation}
For the numerical analysis, we parameterize the dark component through the dark-matter Fermi momentum $k_{F\chi}$. The representative values considered here are as, 
\begin{equation}
k_{F\chi} = 0,\, 20,\, 30,\, 40~{\rm MeV},
\label{eq:dm_fermi_momenta}
\end{equation}
where $k_{F\chi}=0$ corresponds to the reference configuration without dark matter (``No dark matter'' in figures legends).
%
The chemical potential of a baryon species $B$ is defined in terms of its Fermi energy and the corresponding vector mean fields as, 
\begin{equation}
\mu_B = \sqrt{k_{F,B}^2+M_B^{*2}} + g_{\omega B}\omega_0 + g_{\rho B}I_{3B}b_{03}{\color{red}+g_{zB}Z'_0}.
\label{eq:chemical}
\end{equation}
The chemical potential of the dark-matter particle is similarly given by
\begin{equation}
\mu_\chi = \sqrt{k_{F\chi}^2+m_\chi^2} + g_\chi Z'_0.
\label{eq:dm_chemical_potential}
\end{equation}
Next, The total energy density of hyperonic matter in $\beta$ equilibrium, including leptons and dark matter, is expressed as
\begin{align}
\mathcal{E} ={}& \frac{1}{2}\!\left(m_\sigma^2\sigma_0^2 + m_\omega^2\omega_0^2 + m_\rho^2b_{03}^2\right) + 3\Lambda_vg_\omega^2g_\rho^2\omega_0^2b_{03}^2 \nonumber\\
&+ \sum_B\frac{\gamma_B}{2\pi^2}\!\int_0^{k_{F,B}}\!dk\,k^2 \sqrt{k^2+M_B^{*2}} \nonumber \\
&+ \sum_{\ell=e,\mu}\frac{1}{\pi^2}\!\int_0^{k_{F,\ell}}\!dk\,k^2 \sqrt{k^2+m_\ell^2} \nonumber \\
&+ \frac{\gamma_\chi}{2\pi^2}\!\int_0^{k_{F\chi}}\!dk\,k^2 \sqrt{k^2+m_\chi^2} + \frac{1}{2}m_{Z^\prime}^2Z_0^{\prime\,2}.
\label{eq:energy_density}
\end{align}
The pressure is correspondingly given by
\begin{align}
\mathcal{P} =& - \frac{1}{2} m_\sigma^2\sigma_0^2 + \frac{1}{2} m_\omega^2\omega_0^2 + \frac{1}{2} m_\rho^2b_{03}^2 + \Lambda_v g_\omega^2 g_\rho^2 \omega_0^2 b_{03}^2 \nonumber\\
&+ \frac{\gamma}{3(2\pi)^3} \sum_B \int_0^{k_{F,B}} \frac{ k^2\,d^3k}{\sqrt{k^2+M_B^{*2}}} \nonumber\\
&+ \frac{1}{3\pi^2} \sum_l \int_0^{k_{F,l}} \frac{k^4\,dk}{\sqrt{k^2+m_l^2}} \nonumber\\
&+ \frac{\gamma}{3(2\pi)^3} \int_0^{k_{F\chi}} \frac{k^2\,d^3k}{\sqrt{k^2+M_\chi^2}} + \frac{1}{2} m_{Z'}^2 Z_0'^2.
\label{eq:pressure}
\end{align}

The $Z^\prime$ contribution in Eqs.~\eqref{eq:chemical} and \eqref{eq:dm_chemical_potential} is essential for a thermodynamically consistent treatment of the vector portal. It represents the repulsive shift of the single-particle energies induced by the vector mean field. For matter in weak-interaction equilibrium, the chemical potentials satisfy
\begin{equation}
\mu_B = b_B\mu_n - q_B\mu_e,
\label{eq:beta_equilibrium_general}
\end{equation}
where $b_B=1$ is the baryon number and $q_B$ is the electric charge of
species $B$ in units of the proton charge. The charge neutrality condition is
\begin{equation}
\sum_B
q_B\rho_B
+
\sum_{l=e,\mu}
q_l\rho_l
=
0 ,
\label{eq:neutrality}
\end{equation}
where $q_B$ and $q_l$ denote the electric charges of baryons and leptons, respectively. 
The hyperon--meson couplings are parameterized as 
$g_{\sigma B} = x_{\sigma B}g_{\sigma N}$, $g_{\omega B} = x_{\omega B}g_{\omega N}$, and $g_{\rho B} = x_{\rho B}g_{\rho N}$. The strange quark is assumed not to couple directly to the $\sigma$ and $\omega$ mesons, $g_\sigma^s=g_\omega^s=0$, while the nucleon couplings satisfy $g_\omega=3g_\omega^q$, and $g_\rho=g_\rho^q$. The model parameters are fixed by reproducing the empirical saturation properties of nuclear matter, including the saturation density, binding energy per nucleon, and symmetry energy coefficient.

\section{Neutron Star Structure, Tidal Deformability, and Moment of Inertia}
\label{sec:ns_tidal}
Once the equation of state (EOS) of $\beta$-equilibrated hyperonic
matter with vector-portal dark matter has been obtained, it can be
used to determine the macroscopic properties of compact stars. In
this section, we describe the formalism used to calculate the
mass--radius relation, tidal deformability, and moment of inertia.
These observables provide complementary probes of the high-density
behavior of the EOS and, consequently, of the interplay between
hyperonic degrees of freedom and the repulsive interaction generated
by the vector mediator. Once the equation of state (EOS) is obtained, the macroscopic stellar properties are determined by solving the Einstein field equations for a static, spherically symmetric spacetime. 

We consider a static, spherically symmetric configuration for the
background stellar structure. The effects of rotation are neglected
in determining the mass--radius relation and tidal deformability,
while the moment of inertia is calculated subsequently in the
slow-rotation approximation. The EOS obtained in Sec.~\ref{sec:eos}
is used as the microscopic input throughout this analysis.

The spacetime metric describing a non-rotating compact star is given by~\cite{Weinberg:1972kfs}
\begin{equation}
ds^2
=
e^{2\nu(r)}dt^2
-
e^{2\lambda(r)}dr^2
-
r^2
\left(
d\theta^2
+
\sin^2\theta\, d\phi^2
\right),
\label{eq:metric}
\end{equation}
where $\nu(r)$ and $\lambda(r)$ are metric functions depending only on the radial coordinate $r$.

It is convenient to introduce the enclosed gravitational mass function $m(r)$ through the relation
\begin{equation}
e^{2\lambda(r)}
=
\left(
1-\frac{2\,m(r)}{r}
\right)^{-1}.
\label{eq:massfunction}
\end{equation}

\subsection{Tolman--Oppenheimer--Volkoff Equations}
\label{subsec:tov}

The equilibrium structure of compact stars is obtained by solving the Tolman--Oppenheimer--Volkoff (TOV) equations~\cite{Oppenheimer:1939ne,PhysRev.55.364},
\begin{eqnarray}
\frac{d\mathcal{P}(r)}{dr} &=& - \frac{\left[\mathcal{E}(r)+\mathcal{P}(r)\right]\left[m(r)+4\pi r^3 \mathcal{P}(r)\right]}{r\left(r - 2m(r)\right)}, \label{eq:tov_pressure}\nonumber \\
\frac{dm(r)}{dr} &=& 4\pi r^2\mathcal{E}(r). \label{eq:tov_mass}
\end{eqnarray}

Here, $\mathcal{E}(r)$ and $\mathcal{P}(r)$ denote the energy density and pressure profiles inside the star, respectively. In the present work, we adopt natural units with $c=1$ unless explicitly stated otherwise. For a given EOS, the TOV equations are integrated outward from the stellar center using the boundary conditions $m(0)=0$, $\mathcal{P}(0)=\mathcal{P}_c,$. Where $\mathcal{P}_c$ is the central pressure corresponding to a chosen central energy density $\mathcal{E}_c$. The stellar radius $R$ is determined by the condition $\mathcal{P}(R)=0$ while the total gravitational mass of the star is given by $M=m(R)$. The maximum mass obtained from the TOV solutions provides an important constraint on the stiffness of the EOS and its compatibility with recent astrophysical observations.

\subsection{Tidal deformability}
\label{subsec:tidal_deformability}
The tidal response of a neutron star in a compact binary system
provides a direct connection between the microscopic equation of
state (EOS) and the gravitational-wave signal emitted during the
inspiral phase. As the two compact objects approach each other, the
gravitational field of one component induces a quadrupolar deformation
in its companion. The magnitude of this deformation is characterized
by the tidal deformability parameter, which therefore provides an additional probe of the internal structure of neutron stars and the
high-density behavior of dense matter. 
At leading order, the induced quadrupole moment $Q_{ij}$ is related to
the external tidal field $\mathcal{E}_{ij}$ through
\begin{equation}
Q_{ij} = -\lambda \mathcal{E}_{ij},
\label{eq:quadrupole_tidal}
\end{equation}
where $\lambda$ denotes the dimensional tidal deformability
\cite{Hinderer:2007mb, Hinderer:2009ca, Binnington:2009bb}. For the quadrupolar mode with $\ell=2$, the
tidal deformability is related to the corresponding Love number
$k_2$ according to~\cite{Flanagan:2007ix, Hinderer:2009ca, Postnikov:2010yn}
\begin{equation}
k_2 = \frac{3}{2}\frac{\lambda}{R^5},
\label{eq:k2_lambda}
\end{equation}
where $R$ represents the radius of the neutron star. The quadrupolar response can be obtained by considering a linear
perturbation of the spherically symmetric stellar spacetime generated
by the external tidal field. We work in the Regge--Wheeler gauge,
where the metric perturbation associated with the $\ell=2$ mode can be
written as \cite{Hinderer:2007mb, Hinderer:2009ca, Binnington:2009bb}
\begin{eqnarray}
&& h_{\alpha \beta} = \mathrm{Diag} \Big[ -e^{2\nu(r)}H_0(r), e^{2\lambda(r)}H_2(r), r^2K(r),\nonumber \\
&& \hspace{3cm} r^2\sin^2\theta K(r) \Big]Y_{20}(\theta,\phi).
\label{eq:hegge}
\end{eqnarray}
Here, $H_0(r)$, $H_2(r)$, and $K(r)$ denote the radial metric
perturbation functions, while $Y_{20}(\theta,\phi)$ is the
corresponding spherical harmonic. Using the linearized Einstein equations, one obtains
\begin{equation}
H_2(r)=-H_0(r)\equiv H(r),
\end{equation}
together with
\begin{equation}
K'(r)=2H(r)\nu'(r).
\end{equation}
It is therefore sufficient to consider the radial behavior of the
function $H(r)$. Introducing its logarithmic derivative,
\begin{equation}
y(r) = r\frac{H_0'(r)}{H_0(r)}, \label{eq:y_definition}
\end{equation}
the coupled perturbation equations can be reduced to the following
first-order differential equation \cite{Damour:2009vw}:
\begin{equation}
r\,y'(r) + y(r)^2 + y(r)F(r) + r^2Q(r) = 0. \label{eq:tidal_y}
\end{equation}
The functions $F(r)$ and $Q(r)$ appearing in Eq.~\eqref{eq:tidal_y} are determined by the background stellar configuration and the underlying EOS. They are given by
\begin{eqnarray}
F(r) &=& \left[ 1+4\pi r^2 \left( \mathcal{P}-\mathcal{E} \right)\right] \left(1-\frac{2M}{r}\right)^{-1}, \label{fr} \nonumber\\
Q(r) &=& 4\pi \left[ 5\mathcal{E} + 9\mathcal{P} + \frac{\mathcal{E}+\mathcal{P}}{d\mathcal{P}/d\mathcal{E}} \right] \left(1-\frac{2M}{r}\right)^{-1} \nonumber\\
&& - \frac{6}{r^2} \left( 1-\frac{2M}{r} \right)^{-1} \nonumber\\
&& - \frac{4M^2}{r^4} \left( 1+\frac{4\pi r^3\mathcal{P}}{M}\right)^2 \left( 1-\frac{2M}{r} \right)^{-2}. \label{qr}
\end{eqnarray}
Here, $\mathcal{E}$ and $\mathcal{P}$ denote the energy density and pressure of the stellar matter, respectively. The quantity $d\mathcal{P}/d\mathcal{E}$ represents the squared speed of sound $c_s^2$ in the stellar medium and therefore contains direct information about the density dependence of the EOS. For a given EOS, Eq.~\eqref{eq:tidal_y} is solved simultaneously with the TOV equations, Eqs.~\eqref{eq:tov_pressure} and \eqref{eq:tov_mass}, from the stellar center to its surface. The appropriate regularity conditions at the center are as follows
\begin{equation}
y(0)=2, \qquad \mathcal{P}(0)=\mathcal{P}_c, \qquad M(0)=0, 
\end{equation}
where $\mathcal{P}_c$ denotes the central pressure. The integration then yields the surface value
\begin{equation}
y_R \equiv y(R),
\end{equation}
which contains the information about the response of the complete stellar density profile to the external tidal field. The quadrupolar Love number $k_2$ can be expressed in terms of the stellar compactness and the surface value $y_R$ as
\begin{eqnarray}
k_2 &=& \frac{8C^5}{5} (1-2C)^2 \left[ 2+2C(y_R-1)-y_R \right] \nonumber\\
&\times& \bigg\{2C \left[ 6-3y_R+3C(5y_R-8) \right] \nonumber\\
&+& 4C^3 \left[ 13-11y_R + C(3y_R-2) + 2C^2(1+y_R) \right] \nonumber\\
&+& 3(1-2C)^2 \left[ 2-y_R+2C(y_R-1) \right] \ln(1-2C) \bigg\}^{-1}, \nonumber \\
\label{love_number_k2}
\end{eqnarray}
where, $C\equiv\frac{M}{R}$ is the compactness of the star in geometrized units, $G=c=1$. The dimensionless tidal deformability is obtained by normalizing the dimensional quantity $\lambda$ by the fifth power of the stellar mass. It is therefore defined as
\begin{equation}
\Lambda = \frac{\lambda}{M^5} = \frac{2k_2}{3C^5}. \label{eq:tidal}
\end{equation}
The strong dependence of $\Lambda$ on the compactness, $\Lambda\propto C^{-5}$, makes this observable particularly sensitive to changes in the neutron-star radius and, consequently, to modifications of the EOS at supranuclear densities \cite{Flanagan:2007ix, Hinderer:2007mb, Hinderer:2009ca, Damour:2012yf}. In the present work, the tidal deformability is calculated using the  EOS of hyperonic matter obtained within the MQMC framework in the presence of the vector-portal dark-matter interaction. Thus, any change in the composition, pressure, or energy density produced by the appearance of hyperons or by the $Z^\prime$-mediated interaction is propagated self-consistently into the tidal response through $y(r)$, $k_2$, and $\Lambda$. The tidal deformation of each component also leaves a cumulative imprint on the gravitational-wave phase during the inspiral of a binary neutron-star system. Consequently, the gravitational-wave signal does not depend on the tidal deformability of either star independently, but rather on a particular mass-weighted combination of the two individual tidal deformabilities. For a binary with component masses $M_1$ and $M_2$, the effective dimensionless tidal deformability is defined as~\cite{LIGOScientific:2017vwq,Flanagan:2007ix,Hinderer:2009ca,Hinderer:2007mb}
\begin{equation}
\widetilde{\Lambda} = \frac{16}{13}\left[\frac{(M_1+12M_2)M_1^4\Lambda_1 + (M_2+12M_1)M_2^4\Lambda_2}{(M_1+M_2)^5}\right],
\label{eq:tidal-final}
\end{equation}
where $\Lambda_1$ and $\Lambda_2$ denote the individual dimensionless
tidal deformabilities of the two neutron stars \cite{Damour:2012yf}. The quantity $\widetilde{\Lambda}$ is therefore the appropriate
combination for confronting the theoretical EOS with the tidal
contribution to the gravitational-wave phase. Since both
$\Lambda_i$ depend strongly on the corresponding stellar radii and
internal density distributions, measurements of
$\widetilde{\Lambda}$ provide a sensitive probe of the dense-matter
EOS. In the present hyperonic-star model, the tidal observables are
particularly useful for distinguishing the competing effects of
hyperonic softening and vector-mediated repulsion. The appearance of
hyperons modifies the pressure and density profile and generally
reduces the tidal deformability, whereas the repulsive $Z^\prime$
interaction can increase the pressure support and alter the stellar
compactness. Consequently, the variation of $\Lambda$ and
$\widetilde{\Lambda}$ with the dark-matter Fermi momentum
$k_{F\chi}$ provides an additional macroscopic diagnostic of the
vector-portal interaction, complementary to the maximum mass and
radius. 
\subsection{Moment of inertia}
\label{subsec:moment_inertia}
The moment of inertia provides an additional probe of the internal
mass distribution of a neutron star and is particularly sensitive to
the stellar radius and the density profile in the region of
intermediate and high baryon density. The moment of inertia
$I=J/\Omega$, J being the angular momentum and $\Omega$ being the angular frequency measured by distance observer can be calculated using the mass, energy density profile from the  TOV equation. The dimensionless ratio involving the
moment of inertia given as
\begin{equation}
    \frac{I}{MR^2}=\frac{1}{2C}\frac{w_R}{3+w_R},\quad w_R=\frac{r}{\omega}\frac{d\omega}{dr}\Big|_{r=R}
\end{equation}
can be calculated by solving the differential equation for the function $w(r)$, which is related to the space time metric function associated with a steady rotating neutron star, given as\cite{Wei:2018dyy, Lattimer:2015nhk}
\begin{equation}
\frac{dw}{dr} = \frac{4\pi r\left(\mathcal{E}+P\right)\left(4+w\right)}{\left(1-\frac{2m}{r}\right)} - \frac{w(3+w)}{r},
\label{eq:w_equation}
\end{equation}
with the boundary condition $w(0)=0$. The function $w(r)$ is integrated simultaneously with the TOV equations from the center to the stellar surface. The resulting moment of inertia therefore depends on the complete internal density and pressure profiles rather than only on the total mass and radius. The stellar observables calculated in this section are determined entirely by the EOS constructed in Sec.~\ref{sec:eos}. 

The influence
of vector-portal dark matter therefore enters the stellar structure
through the modified pressure and energy density.
The vector mediator generates an additional repulsive contribution
to the high-density matter sector. Increasing the dark-matter
admixture therefore modifies the pressure support against
gravitational collapse and can change the maximum mass and radius of
the star. At the same time, the appearance of hyperons generally
softens the EOS by opening additional degrees of freedom. The
observable stellar properties consequently reflect the competition
between these two effects. 
For the present analysis, we solve the TOV equations for the EOSs
corresponding to
\begin{equation}
k_{F\chi}
=
0,\;20,\;30,\;40~{\rm MeV},
\label{eq:kfchi_star}
\end{equation}
and investigate the resulting changes in
$M_{\rm max}$, $R$, $\Lambda$, and $I$. In addition, the dependence
on the nonlinear coupling $\Lambda_v$ and the hyperon optical
potential, particularly $U_\Xi$, is examined. 
Thus, the simultaneous determination of the mass--radius relation,
tidal deformability, and moment of inertia provides a complementary
set of constraints on the dense-matter EOS and allows us to quantify
the impact of vector-portal dark matter in hyperonic compact stars.

\section{RESULTS AND DISCUSSION}
\label{sec:results}
We first discuss the parameters used in MQMC without introducing vector portal dark matter. These are constrained using the properties of free
baryons and symmetric nuclear matter. The MQMC model parameters contain  the harmonic-oscillator confinement strength
$a$, the constant potential parameter $V_0$ and the constituent quark masses. In the present calculation, the constituent-quark masses are taken as 
$m_u=m_d=150~{\rm MeV}$, 
$m_s=250~{\rm MeV}$, 
with 
$a=0.69655~{\rm fm}^{-3}$. 
The value of $V_0$ for each member of the baryon octet is determined
by reproducing the corresponding free-space baryon mass. The resulting
values are summarized in Table~\ref{tab:V0_parameters}.

\begin{table}[h!]
\centering
\caption{Free-space baryon masses and corresponding MQMC confinement
parameter $V_0$ for $a=0.69655~{\rm fm}^{-3}$,
$m_u=m_d=150~{\rm MeV}$, and $m_s=250~{\rm MeV}$.}
\label{tab:V0_parameters}
\begin{tabular}{ccc}
\hline
Baryon & $M_B$ (MeV) & $V_0$ (MeV) \\
\hline
$N$       & 939.0  & 44.05  \\
$\Lambda$ & 1115.6 & 69.46  \\
$\Sigma$  & 1193.1 & 86.47  \\
$\Xi$     & 1321.3 & 100.06 \\
\hline
\end{tabular}
\end{table}
The quark--meson coupling constants are determined by reproducing the
empirical properties of symmetric nuclear matter. In particular, the
saturation density and binding energy per nucleon are fixed at
\begin{equation}
\rho_0=0.15~{\rm fm}^{-3},
\qquad
\frac{\mathcal{E}}{\rho_B}-M_N=-15.7~{\rm MeV},
\end{equation}
Together with condition $P(\rho_0)=0$ and symmetry-energy 
coefficient $J=32~{\rm MeV}$. 
The meson masses used in the calculation are $m_\sigma=550~{\rm MeV}$, 
$m_\omega=783~{\rm MeV}$ and $m_\rho=763~{\rm MeV}$. For $m_q=150~{\rm MeV}$, the reference parameter set without dark
matter is $g_\sigma^q=4.39952$, $g_\omega=6.74299$, $g_\rho=9.78255$ and $\Lambda_v=0.1$. 

Next we discuss the parameters for the DM sector with the vector portal. 
The interaction between the dark and baryonic sectors is mediated by a massive vector boson $Z^\prime$, with the strength of the baryonic coupling determined by $g_{BZ^\prime}\equiv 3g_q$, while $g_{\chi Z^\prime}$ denotes the coupling of the mediator to the dark-matter fermion \cite{Bishara:2017pfq,Borah:2025cqj}. In an earlier study
\cite{Kumar:2026hoq}, it was found that for a relatively heavy dark
matter particle and mediator, with $m_\chi$ and $M_{Z^\prime}$ of
order $1~{\rm TeV}$, the contribution of the vector portal to the
neutron-star equation of state is rather small. Such a parameter
region is also associated with the limiting cases allowed by the
combined requirements from the relic abundance, direct-detection
searches, and collider constraints, including dijet searches
\cite{Taramati:2024kkn}. Since our primary objective here is to
investigate the possible impact of the dark sector on the EOS and
global properties of hyperonic stars, we instead consider a light
mediator scenario motivated by self-interacting dark matter (SIDM)
\cite{Spergel:1999mh,Tulin:2013teo,Patel:2022qvv,Kouvaris:2014uoa,
Bernal:2015ova,Kainulainen:2015sva,Hambye:2019tjt,Cirelli:2016rnw,
Kahlhoefer:2017umn,Dutta:2021wbn}. Specifically, we take the dark
matter mass to be $m_\chi=10~{\rm GeV}$ and consider a vector
mediator with a mass of order $m_{Z^\prime}\sim100~{\rm MeV}$. This
choice allows the dark sector to possess appreciable self-interactions
while keeping its coupling to the visible sector comparatively
small. Following the coupling hierarchy commonly employed in SIDM
realizations, we assume a relatively strong interaction between
$\chi$ and $Z^\prime$, with $g_{\chi Z^\prime}=\mathcal{O}(0.4)$,
whereas the coupling to quarks is taken in the much smaller range
$g_{qZ^\prime}\sim10^{-5}$--$10^{-3}$. For the benchmark calculation
presented here, we have taken, specifically,
$m_\chi=10~{\rm GeV}, 
m_{Z^\prime}=100~{\rm MeV}, 
g_{\chi Z^\prime}=0.4, 
g_{qZ^\prime}=5\times10^{-4}$, 
with the corresponding baryonic coupling given by $g_{BZ^\prime}=3g_{qZ^\prime}$. This hierarchy between the dark- and
visible-sector couplings provides a useful framework for studying the
effect of a light vector mediator in dense hyperonic matter while
suppressing potentially strong laboratory constraints on its
coupling to ordinary particles. The considered region of parameter
space has also been extensively explored in SIDM studies in the
context of small-scale structure problems associated with the
collisionless cold-dark-matter paradigm, including the core--cusp
problem~\cite{Navarro:1996gj}, the too-big-to-fail problem
\cite{Boylan-Kolchin:2011qkt,Boylan-Kolchin:2011lmk}, and the
missing-satellites problem
\cite{Klypin:1999uc,Kauffmann:1993gv,Moore:1999nt}. In particular,
the presence of a light mediator can generate velocity-dependent
dark-matter self-scattering, thereby allowing sizeable
self-interaction rates on galactic scales while remaining compatible
with constraints from larger-scale structures and galaxy clusters
\cite{Bahcall:1999xn,Springel:2006vs,Trujillo-Gomez:2010jbn}. The
suppressed coupling to quarks also reduces the strength of the
corresponding direct-detection and collider signatures, while the
observed relic abundance can be accommodated through annihilation
processes within the dark sector involving the light mediator
\cite{Tulin:2013teo,Borah:2021pet,Patel:2022qvv}. We therefore use
the parameter set summarized in Table~\ref{tab:quark_couplings} as a
representative benchmark for examining how a light vector mediator
modifies the thermodynamic properties of $\beta$-equilibrated
hyperonic matter and, subsequently, the mass--radius relation, tidal
deformability, and moment of inertia of dark-matter-admixed
hyperonic stars.\\
The dark-matter Fermi momentum $k_{F\chi}$ is a key parameter
controlling the impact of the vector portal on hyperonic stellar
matter. For degenerate fermionic dark matter,
$n_\chi\propto k_{F\chi}^{3}$, and hence the source of the portal
field increases rapidly with $k_{F\chi}$,
\begin{equation}
Z'_0 \propto g_{\chi Z'}n_\chi
\propto g_{\chi Z'}k_{F\chi}^{3}.
\end{equation}
Therefore, a sufficiently large dark-matter density can generate a
non-negligible $Z'_0$ mean field, which modifies the baryonic chemical
potentials and may eventually influence the saturation properties of  hyperonic matter. This requires refitting of the RMF parameters. In the present analysis, for each $k_{F\chi}$, we refit the coupling parameters so that nuclear saturation properties are satisfied as presented in Table~\ref{tab:quark_couplings}.  
The dependence on $k_{F\chi}$ is consequently propagated to the macroscopic stellar observables, particularly $M_{\rm max}$, $R$, $\Lambda$, and $I$. 

\begin{table}[h!]
\centering
\caption{MQMC quark--meson coupling parameters used for
$m_q=150~{\rm MeV}$ for different dark-matter fermi momenta.
The couplings are those employed in the numerical calculations. We have taken the coupling $g_{BZ'}=0.00001$ and $g_{\chi Z'}=0.4$.}
\label{tab:quark_couplings}
\begin{tabular}{ccccc}
\hline
$k_{F\chi}$ (MeV)
& $g_\sigma^q$
& $g_\omega$
& $\Lambda_v$
& $g_\rho$
\\
\hline
$0$  & 4.39952 & 6.74299 & 0.1 & 9.78255 \\
$20$ & 4.53961 & 6.85160 & 0.1 & 9.83292 \\
$30$& 4.84859 & 7.04736 & 0.1 & 9.92591 \\
$40$& 5.38866 & 7.25852 & 0.1 & 10.01074\\
\hline
\end{tabular}
\end{table}
For the hyperon sector, the hyperon--meson couplings play a crucial role in determining the composition and stiffness of dense hyperonic matter. In the present MQMC calculation, the scalar response of each baryon is obtained self-consistently from its effective mass, while the vector
couplings are constrained by the hyperon optical potentials in
symmetric nuclear matter.  We adopt the reference values $U_\Lambda=-28~{\rm MeV}$, $U_\Sigma=+30~{\rm MeV}$ and $U_\Xi=-18~{\rm MeV}$. For different values of the dark matter fermi momenta, the corresponding refitted hyperon--meson coupling ratios are presented in Table~\ref{tab:hyperon_couplings}. The $\Lambda$ hyperon potential is constrained from single-particle levels in $\Lambda$ hypernuclei over a broad mass range~\cite{Millener1988}. Experimental analyses of $\Sigma^-$ atomic data indicate a repulsive $\Sigma$--nucleus interaction in the nuclear interior~\cite{Mares1995,Bart1999} as $U_\Xi \sim -(14-18)~{\rm MeV}$. 

\begin{table}[t]
\centering
\caption{The fitted hyperon--meson coupling ratios
$x_{\omega B}=g_{\omega B}/g_{\omega N}$ for the hyperon potentials taken here for different values of dark matter fermi momenta. We have taken the coupling $g_{BZ'}=0.00001$ and $g_{\chi Z'}=0.4$.}
\label{tab:hyperon_couplings}
\begin{tabular}{cccc}
\hline
$k_{F\chi}$ (MeV)
& $x_{\omega\Lambda}$
& $x_{\omega\Sigma}$
& $x_{\omega\Xi}$
\\
\hline
$0$  & 0.81  & 1.59 & 0.246 \\
$20$ & 0.825 & 1.58 & 0.25  \\
$30$ & 0.84  & 1.59 & 0.27  \\
$40$ & 0.935 &1.62 &0.29\\
\hline
\end{tabular}
\end{table}

\begin{figure}
\centering
\includegraphics[width=0.9\linewidth]{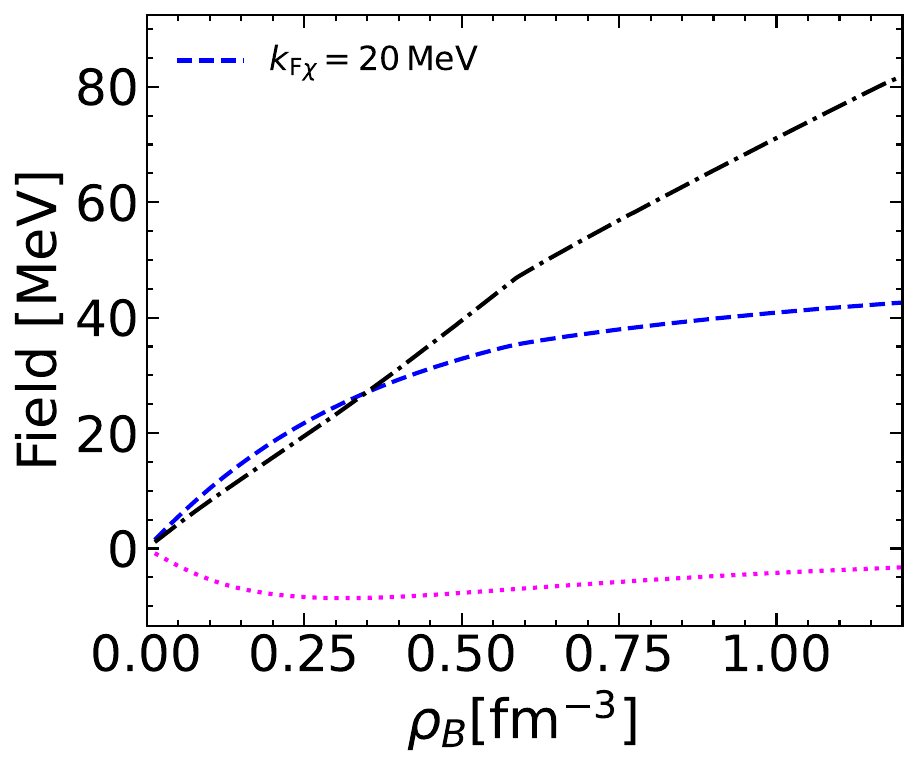}
\includegraphics[width=0.95\linewidth]{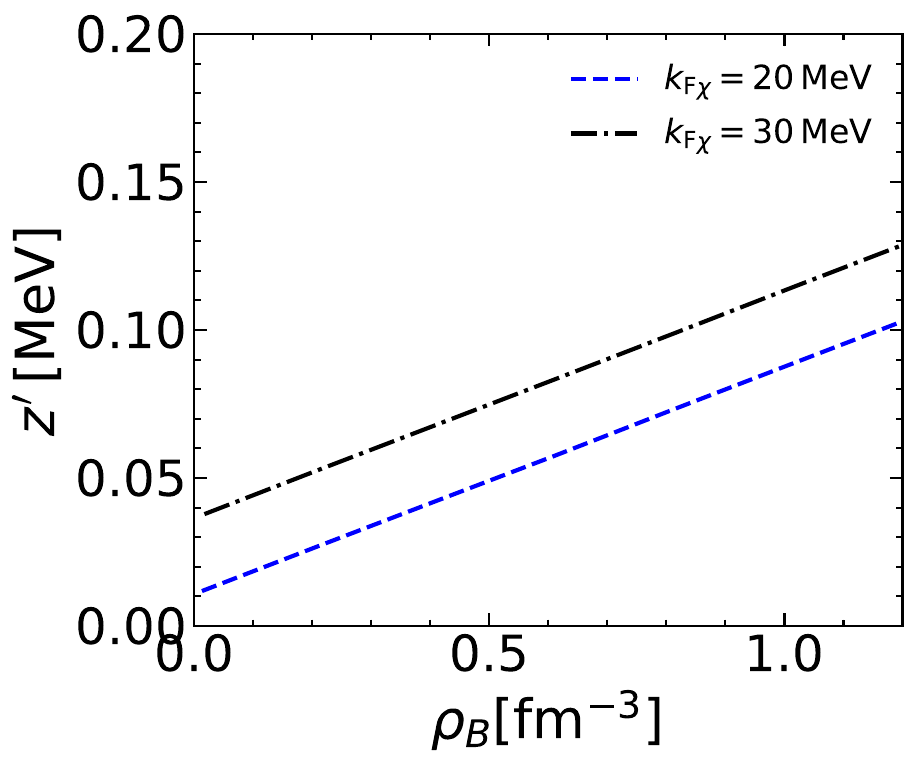}
\includegraphics[width=0.95\linewidth]{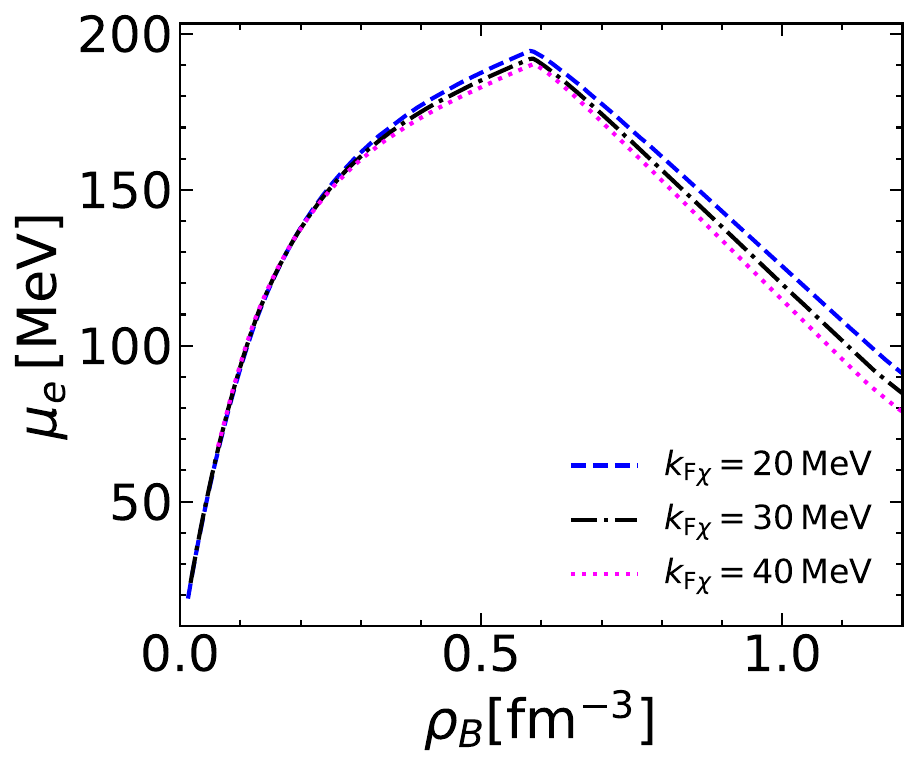}
\caption{Density dependence of the self-consistent mean fields and
chemical potentials. Left: Variation of the mean meson fields ($\sigma$, $\omega$, and $\rho$) as a function of baryon density, $\rho_B$, at fixed dark matter Fermi momenta, $k^{F\chi} = 20$ MeV. Middle: temporal component of the vector mediator, $Z'_0$, for
$k_{F\chi}=20$ and $30~{\rm MeV}$. Right: electron chemical potential
$\mu_e$ for $k_{F\chi}=0$, $20$, and $30~{\rm MeV}$. The results illustrate the influence of the dark matter Fermi momentum on the self-consistent meson mean fields. \label{fig:fields}}
\end{figure}
\subsection{Equation of state and the effect of vector-portal
dark matter}
\label{subsec:results_eos}
\begin{figure}[htb!]
\centering
\includegraphics[width=0.95\linewidth]{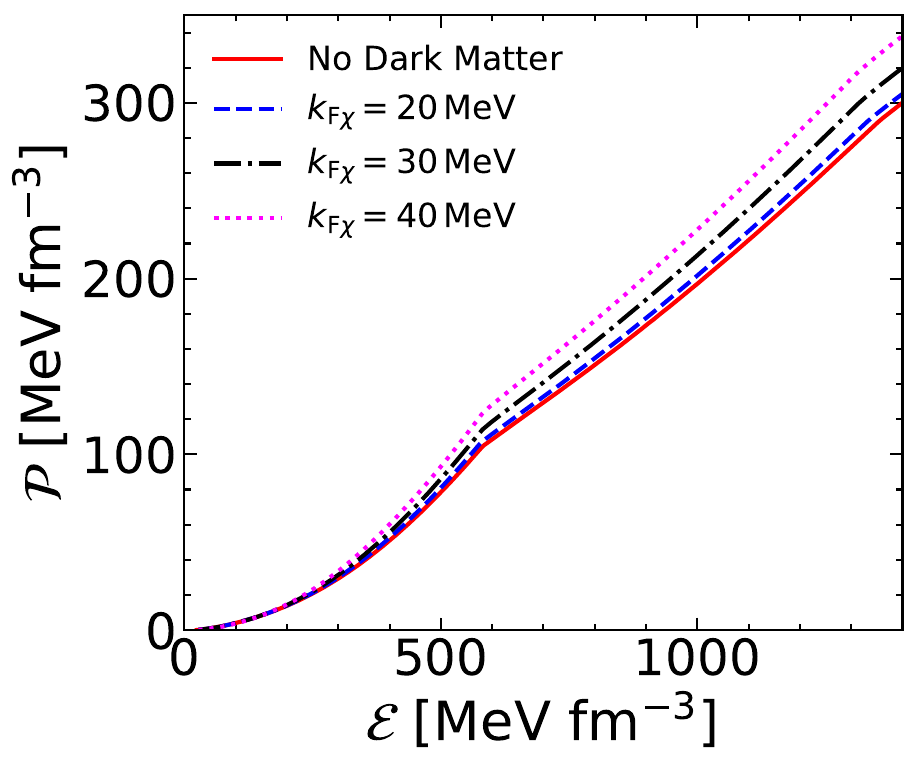}
\caption{Equation of state for different dark-matter Fermi momenta, $k_{F\chi}=20$, $30$, and $40~{\rm MeV}$.}
\label{fig:eos}
\end{figure}

Next we discuss EOS for the hyperonic matter including the cases without dark matter and with different values of the dark-matter Fermi momentum. Figure \ref{fig:eos} illustrates the equation of state (EOS), expressed as the pressure $P$ as a function of the energy density $\varepsilon$, for several fixed DM Fermi momenta, $k_{F\chi} = 20,\, 30,\, 40$ MeV and without DM. The solid red curve denotes the hyperonic EOS without dark matter while the other lines correspond to different values of $k_{F\chi}$. The inclusion of hyperons softens the EOS
relative to purely nucleonic matter. This behavior originates from
the appearance of additional baryonic degrees of freedom, which
redistribute the baryon density among several species and reduce the
Fermi pressure carried by the nucleonic component. 

The inclusion of vector-portal dark matter produces an additional
modification of the EOS. As $k_{F\chi}$ increases, the contribution
of the dark sector to the total energy density and pressure increases.
At the same time, the $Z^\prime$ mean field modifies the
self-consistent baryonic mean fields and chemical potentials.
Consequently, the EOS becomes stiffer. Increasing $k_{F\chi}$ alters the balance between the attractive scalar and repulsive vector interactions, leading to a noticeable change in the stiffness of the EOS which supports a larger maximum mass. These results demonstrate that the dark matter Fermi momentum serves as an important parameter governing the thermodynamic properties of dense matter and the macroscopic structure of compact stars.

The influence of the vector portal is also visible directly in the
density dependence of the self-consistent mean fields. Figure~\ref{fig:fields}
shows the $\sigma$, $\omega$, and $\rho$ fields as functions of the
baryon density for a representative value
$k_{F\chi}=20~{\rm MeV}$. The same figure also displays the
corresponding $Z'_0$ field for different dark matter fermi momenta $k_{F\chi}=20$ and $30~{\rm MeV}$.  The scalar mean field provides the attractive contribution that lowers
the effective baryon masses, whereas the $\omega$ field generates
the dominant repulsive interaction at high density. The $\rho$ field
controls the isovector sector and is therefore particularly important
for the proton fraction and the onset of negatively charged hyperons.

The $Z'_0$ field increases with the dark-matter Fermi momentum because
the vector mediator is sourced by both the baryonic and dark-matter
vector densities. Thus,
\begin{equation}
Z'_0
=
\frac{
g_{\chi Z'}\rho_\chi+\sum_B g_{BZ^\prime}\rho_B
}{
m_{Z^\prime}^2
},
\end{equation}
and a larger dark-matter density results in a stronger vector portal 
mean field. 
\subsection{Mass--radius relation and maximum mass}
\label{subsec:results_mass_radius}
\begin{figure}[htb!]
\centering
\includegraphics[width=0.9\linewidth]{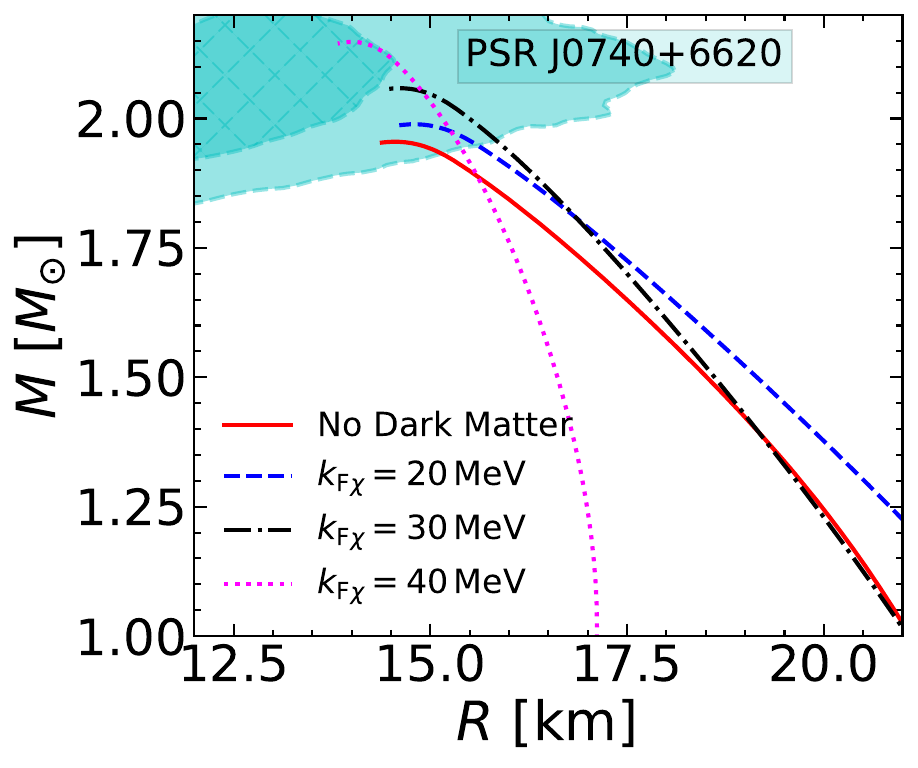}
\caption{Mass--radius relations of hyperonic neutron stars for different dark-matter Fermi momenta, $k_{F\chi}=20$, $30$, and $40~{\rm MeV}$ and without DM (No Dark matter). The shaded regions correspond to the astrophysical observation of the largest NS, PSR J0740 + 6620, observed until now, i.e., of mass $2.08 \pm 0.07\, M_{\odot}$ \cite{Dittmann:2024mbo}. The dark shaded cyan is for  $2\sigma$ while a light shaded cyan is at $3\sigma$ confidence level.}
\label{fig:mass-radius}
\end{figure}

We next discuss the mass--radius relation of
hyperonic neutron stars in the presence of vector-portal dark matter. 
Figure~\ref{fig:mass-radius}
shows the resulting $M-R$ sequences for the reference hyperonic
matter configuration without dark matter and for different values of
the dark-matter Fermi momentum, $k_{F\chi}$. We have plotted alongside also the observational results for the largest NS mass till now i.e., with mass $M_{\rm max}\simeq 2.08\pm 0.07\,M_\odot$ corresponding to  PSR J0740 + 6620~\cite{Dittmann:2024mbo}. The dark cyan band with dotted boundary corresponds to $2\sigma$ while the light cyan band with dashed outline represents $3-\sigma$ confidence interval. The red solid curve in Fig.~\ref{fig:mass-radius} represents M--R curve  in the absence of dark matter denoted as "no DM". The MQMC model with the adopted hyperon interactions can support a maximal mass, $M_{\rm max}\simeq 1.95\,M_\odot$ with radius $R\simeq 14.5~{\rm km}$ even after the softening of the EOS caused by the appearance of strange baryons. 

The inclusion of the vector-portal dark matter produces a noticeable
modification of the mass--radius relation. The blue dashed curve,
corresponding to $k_{F\chi}=20~{\rm MeV}$, lies close to the
no-dark-matter sequence over a substantial part of the mass range,
but extends toward slightly larger stellar masses at the high-density
end. The maximum mass increases to approximately
$M_{\rm max}\simeq 1.99\,M_\odot$ with a corresponding radius $R\simeq 14.8~{\rm km}$. For $k_{F\chi}=30~{\rm MeV}$, represented by the black dash-dotted
curve, the effect of the vector portal becomes more pronounced. The
mass--radius sequence is shifted toward larger masses in comparison
with the no-dark-matter configuration, and the maximum supported mass
increases to approximately
$M_{\rm max}\simeq 2.06\,M_\odot$. The corresponding maximum-mass
radius remains close to $R\simeq 14.61~{\rm km}$. The $k_{F\chi}=40~{\rm MeV}$ configuration, shown by the magenta
dotted curve, exhibits a more distinct behavior. The curve reaches maximum mass
$M_{\rm max}\simeq 2.15\,M_\odot$ with a much shorter radius $R\simeq 14.01~{\rm km}$. This is arising due to additional repulsive contribution generated by the $Z^\prime$ mean field at larger dark-matter density. The dark sector modifies the density dependence of the EOS and consequently changes both the pressure support and the compactness of the star in a mass-dependent manner.

The systematic displacement of the high-mass branches with increasing
$k_{F\chi}$ can be understood from the density dependence of the
vector-portal contribution. For degenerate fermionic dark matter, the
dark-matter number density scales as
\begin{equation}
n_\chi
=
\frac{k_{F\chi}^3}{3\pi^2},
\end{equation}
and therefore the temporal component of the vector mediator is
enhanced as
\begin{equation}
Z'_0
\propto
g_{\chi Z^\prime}n_\chi
\propto
g_{\chi Z^\prime}k_{F\chi}^3.
\end{equation}
Consequently, increasing $k_{F\chi}$ strengthens the vector-mediated
interaction in the dense stellar core. This modifies the baryonic
chemical potentials and the pressure of the hyperonic matter and
provides an additional source of repulsive pressure support against
gravitational collapse.

\begin{table}[htbp]
\centering
\caption{Maximum-mass configurations for $m_q=150~{\rm MeV}$ for
different dark-matter Fermi momenta.}
\begin{tabular}{lccc}
\hline
Configuration
& $\epsilon_c$ (fm$^{-4}$)
& $M_{\rm max}$ ($M_\odot$)
& $R$ (km)
\\
\hline
No Dark Matter & 863.21 & 1.96 & 14.55  \\
$k_{{\rm F}\chi} = 20\, {\rm MeV}$ & 855.31 & 1.99 & 14.80  \\
$k_{{\rm F}\chi} = 30\, {\rm MeV}$ & 851.32 & 2.06 & 14.61 \\
$k_{{\rm F}\chi} = 40\, {\rm MeV}$ & 839.19 & 2.15 & 14.01  \\
\hline
\end{tabular}%
\label{tab:maxmass_dm}%
\end{table}%

The physical effect observed in Fig.~\ref{fig:mass-radius} is
therefore the result of a competition between two mechanisms. The
appearance of hyperons introduces additional degrees of freedom and
generally softens the EOS, reducing the maximum mass that can be
supported by the star. In contrast, the vector-portal interaction
provides an additional repulsive contribution that becomes more
important as the dark-matter Fermi momentum is increased. The shaded cyan region in Fig.~\ref{fig:mass-radius} represents the
observational mass--radius constraint associated with
PSR~J0740+6620~\cite{Dittmann:2024mbo}. The comparison with this region is particularly
useful because the observed object lies in the high-mass regime where
the effects of hyperons and the vector portal are expected to be most
pronounced. Thus, the mass--radius relation provides the first direct macroscopic 
signature of the vector portal in the present hyperonic-star model.  
The increase of the maximum supported mass with increasing
$k_{F\chi}$ demonstrates that the dark sector can partially
counteract the softening produced by hyperonic degrees of freedom. 
The corresponding changes in the radius and compactness are expected
to translate directly into the tidal deformability and moment of
inertia, which are investigated in the subsequent discussions.
\subsection{Tidal Love number, dimensionless tidal deformability}
\label{subsec:results_tidal}
We next calculate the
quadrupolar Love number $k_2$ and the dimensionless tidal deformability
$\Lambda$ using Eqs.~(\ref{love_number_k2}) and (\ref{eq:tidal}) for
the EOSs corresponding to the no-dark-matter case and the different
values of the dark-matter Fermi momentum.
\begin{figure}[htb!]
\centering
\includegraphics[width=0.9\linewidth]{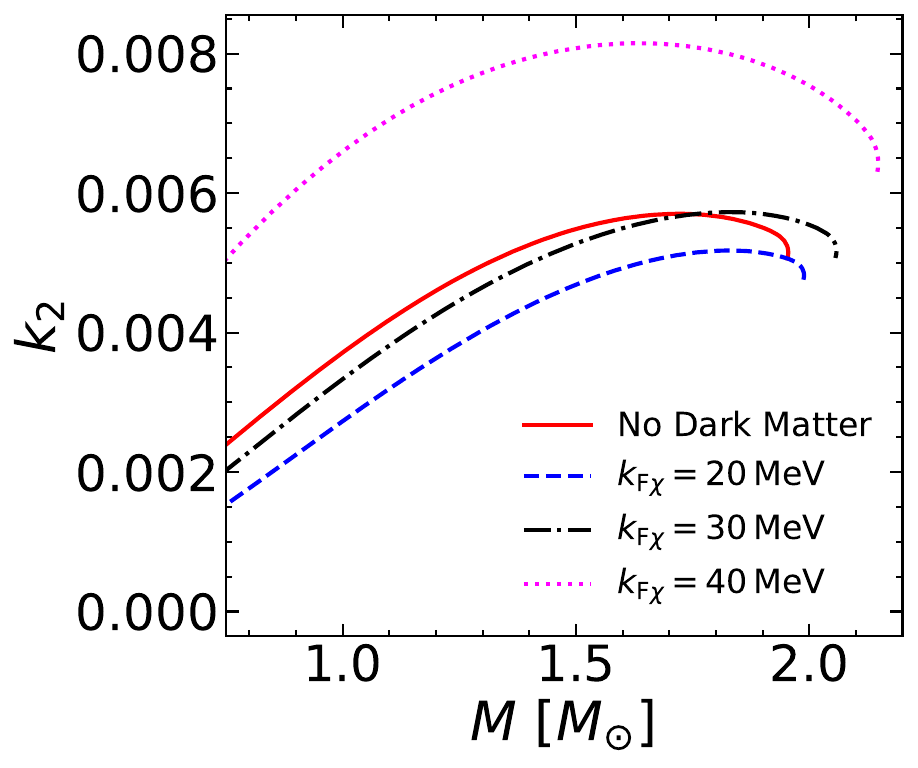}

\vspace{0.2cm}

\includegraphics[width=0.9\linewidth]{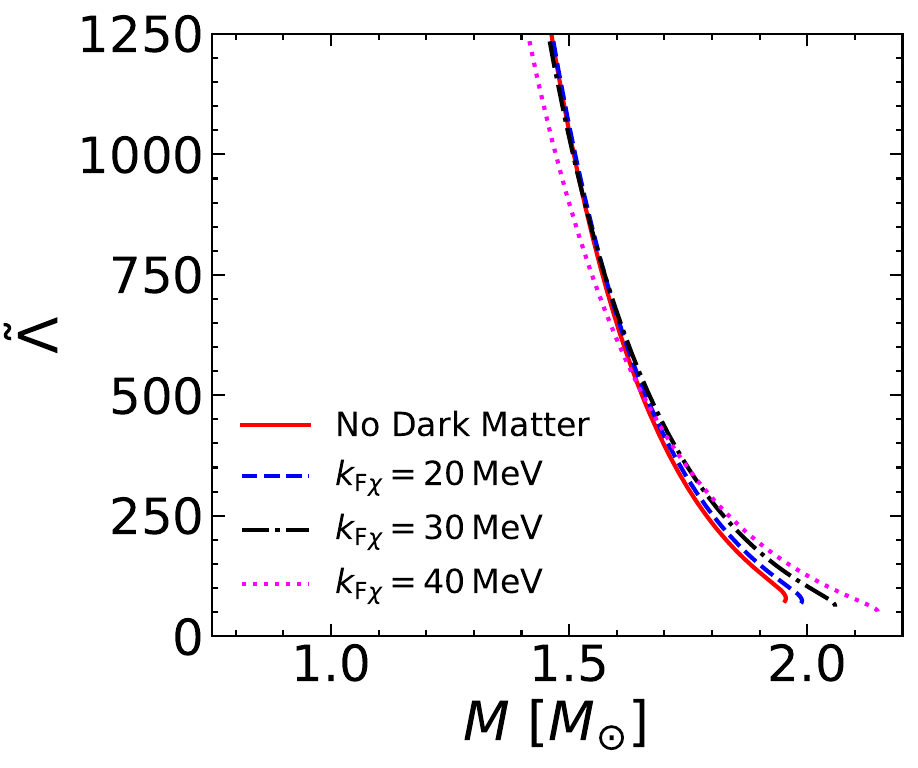}
\caption{Tidal Love number $k_2$ and dimensionless tidal deformability as functions of neutron star mass for different EOS parameterizations considered in the present work.
}
\label{fig:tidal}
\end{figure}
The upper panel of Fig.~\ref{fig:tidal} displays the mass dependence
of the quadrupolar Love number $k_2$. The corresponding dimensionless tidal deformability decreases
rapidly with increasing stellar mass as shown in the lower panel of Fig.~\ref{fig:tidal}. In contrast to the monotonic
behavior of the tidal deformability, the Love number initially
increases with stellar mass and subsequently reaches a maximum before
decreasing toward the high-mass end of each stellar sequence. This behavior is expected
because $\Lambda$ depends strongly on the compactness through
$\Lambda\propto k_2/C^5$ as given in Eq.(\ref{eq:tidal}).

It is to be noted that the inclusion of vector-portal dark matter modifies both the internal density profile and the radius of the star. Consequently, changes in $\Lambda$ arise from two related effects: the direct modification of
the EOS and the resulting change in stellar compactness. 
The sensitivity of $\Lambda$ to the dark-matter Fermi momentum makes
tidal deformability a useful complementary observable to the maximum
mass. Two EOSs that produce similar maximum masses can nevertheless
have different tidal responses because their pressure-density
relations and internal compositions differ at intermediate densities.

\subsection{Moment of inertia}
\label{subsec:moment}
\begin{figure}[htb!]
\centering
\includegraphics[width=0.9\linewidth]{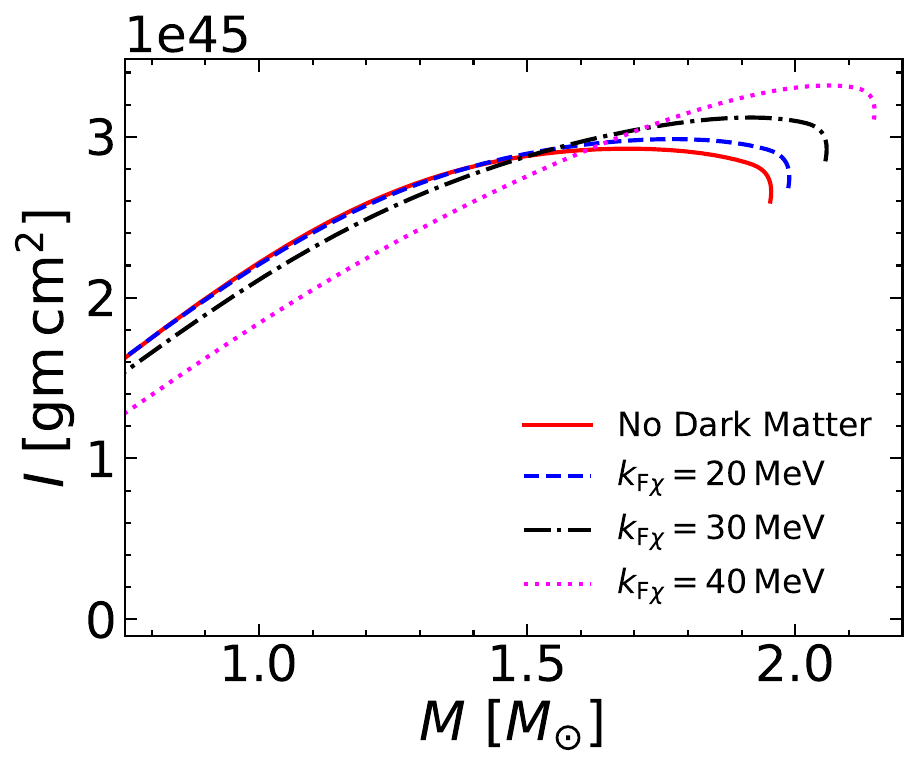}
\caption{Moment of inertia $I$ as a function of the gravitational
mass $M$ for hyperonic neutron stars within the MQMC framework in the
absence of dark matter and for different values of the dark-matter
Fermi momentum, $k_{F\chi}=20$, $30$, and $40~{\rm MeV}$. The moment
of inertia is expressed in units of ${\rm g\,cm^2}$.}
\label{fig:moment_inertia}
\end{figure}

The moment of inertia depends on both the stellar mass and radius as well as on the radial distribution of matter. The estimated value of the moment of inertia $I$ is sensitive to changes in the high-density EOS induced by hyperonic degrees of freedom and the vector-portal dark-matter component. The variation of the moment of inertia with stellar mass absence of dark matter and for different values of the dark-matter
Fermi momentum, $k_{F\chi}=20$, $30$, and $40~{\rm MeV}$ is displayed in Fig.\ref{fig:moment_inertia}. For the configuration without dark matter, $I$ increases from about $1.5\times10^{45}~{\rm g\,cm^2}$ at $M\simeq0.6\,M_\odot$ and reaches a maximum of approximately
$2.9\times10^{45}~{\rm g\,cm^2}$ at a mass close to $1.6\,M_\odot$. Beyond this point, the moment of inertia decreases along the high-mass part of the sequence as the stellar radius decreases toward the maximum-mass configuration. 

The inclusion of the vector-portal dark-matter component changes the
behavior of $I(M)$ in a systematic but nontrivial manner. For
$k_{F\chi}=20~{\rm MeV}$, the curve remains close to the
no-dark-matter sequence over a substantial mass range, but its
high-mass branch extends to slightly larger masses and reaches a
moment of inertia of approximately
$3.0\times10^{45}~{\rm g\,cm^2}$ before turning over. For
$k_{F\chi}=30~{\rm MeV}$, the high-mass sequence is shifted further
toward larger masses, with the maximum moment of inertia approaching
$3.1\times10^{45}~{\rm g\,cm^2}$. The $k_{F\chi}=40~{\rm MeV}$
configuration exhibits the most pronounced modification: although
its moment of inertia is smaller than the other configurations at
lower stellar masses, it increases more rapidly at higher masses and
reaches a maximum of approximately
$3.3\times10^{45}~{\rm g\,cm^2}$.

An interesting feature of Fig.~\ref{fig:moment_inertia} is the
crossing of the different $I(M)$ sequences. At relatively low and
intermediate masses, the $k_{F\chi}=40~{\rm MeV}$ configuration has a
smaller moment of inertia than the corresponding no-dark-matter and
lower-$k_{F\chi}$ configurations. With increasing mass, however,
the curves approach one another and subsequently cross, after which
the larger dark-matter configurations exhibit larger moments of
inertia. This behavior indicates that the effect of the vector portal
cannot be characterized simply as an overall increase in the moment
of inertia. Instead, the dark-matter contribution modifies the
pressure-density relation and consequently the mass distribution and
radius of the star in a density-dependent manner.

At a fixed or comparable stellar mass, an increase in the pressure
support can modify the stellar radius and hence the moment of
inertia. At the same time, the self-consistent $Z^\prime$ mean field
changes the chemical potentials and equilibrium composition of the
hyperonic matter. Therefore, the resulting change in $I$ reflects the
combined effects of the dark-matter contribution, the vector-mediated
repulsion, the hyperonic composition, and the corresponding
modification of the stellar density profile. The high-mass behavior shown in Fig.~\ref{fig:moment_inertia} is
particularly relevant for the present study. As the dark-matter Fermi
momentum is increased from $20$ to $40~{\rm MeV}$, the resulting 
mass of the stellar sequence shifts toward larger values. This is
consistent with the stiffening of the EOS produced by the additional
repulsive contribution associated with the vector portal. The
corresponding increase in the maximum attainable moment of inertia
follows from the larger mass supported by the modified EOS, although
the precise value of $I$ remains sensitive to the stellar radius and
internal density distribution.

\section{Conclusions}
\label{sec:conclusion}
In this work, we have investigated the structure and macroscopic
properties of hyperonic neutron stars in the presence of fermionic
dark matter interacting with baryonic matter through a vector
mediator, $Z^\prime_\mu$, within the modified quark--meson coupling
(MQMC) framework. Then we explored how the additional repulsive interaction generated by the
vector portal competes with the softening of the high-density
equation of state (EOS) caused by the appearance of hyperons. We have
combined the microscopic MQMC description of the baryon structure
with the self-consistent mean-field treatment of the vector portal
and subsequently calculated the global stellar properties by solving
the Tolman--Oppenheimer--Volkoff equations. 

In the absence of dark matter, the baryon–baryon interaction is generated self-
consistently through the coupling of quarks to the $\sigma$,
$\omega$ and $\rho$ mesons in the mean-field approximation. The
model parameters are constrained to reproduce the empirical saturation properties of nuclear matter, including
the binding energy, saturation density, incompressibility,
and symmetry energy. The resulting values of the nuclear
matter incompressibility and the slope parameter of the
symmetry energy are found to be consistent with current
experimental and astrophysical constraints. 

The central result of the present
investigation is that the vector-portal dark
matter can compensate for the hyperon-induced softening of the
EOS depending upon the parameters of the dark sector. The $Z^\prime$ mediator generates an additional vector mean
field, which modified the baryonic chemical potentials and provides
additional repulsive pressure at high density. The strength of this
effect is  controlled by the dark-matter Fermi momentum,
$k_{F\chi}$. Since the number density of degenerate fermionic dark matter is proportional to $k_{F\chi}^3$, increasing
$k_{F\chi}$ enhanced the source of the vector mean field and thereby
modified the high-density pressure and energy density.

The inclusion of the vector-portal dark matter produced a noticeable modification of the mass--radius relation. 
With the parameters of MQMC model with hyperons and without DM, the maximum mass  of the hyperonic NS could 
become as high as $1.96\,M_\odot$ but with a larger radius. Inclusion of vector portal DM consistently increased 
the maximum mass and with sufficiently large density of DM, both the maximum mass and the radius could be within the astrophysical o bservation limits of high mass NS of  PSR J0740+6620~\cite{Dittmann:2024mbo}. It may be mentioned here that while including the DM, 
the MQMC coupling parameters were consistently refitted for each  values of $k_{F\chi}$, so that the saturation 
properties of nuclear matter are reproduced. 

The tidal properties provided an independent characterization of the
modified stellar structure. The quadrupolar Love number $k_2$ and the
dimensionless tidal deformability $\Lambda$ were found to depend
sensitively on the compactness and internal density distribution. 
The different dark-matter configurations produced distinct tidal
responses through their modifications of the EOS and stellar radii.
In particular, the larger-$k_{F\chi}$ configurations exhibited a
more pronounced modification of the Love number and tidal
deformability in the high-mass region. 

The moment of inertia provided another independent probe of the
stellar structure. The calculated $I(M)$ relations
were modified by the dark sector through the corresponding changes
in the EOS, stellar radius, and radial mass distribution. Since the
moment of inertia approximately followed $I\sim M\,R^2$ the changes in radius and mass produced measurable differences in
the moment of inertia among the different dark-matter
configurations. The results therefore complemented the
mass--radius and tidal-deformability predictions and provided an
additional probe of the density-dependent effects of the vector
portal.

In summary, the results demonstrated that while massive hyperonic neutron
stars could have been supported within the MQMC framework despite
the softening associated with the appearance of strange baryons, their raddi are much too large to be within the observational limits. The
additional repulsion generated by the $Z^\prime$ vector portal
provides an interesting  mechanism to compensate  this softening and
allowed the maximum mass to reach the two-solar-mass regime and the corresponding raddi for viable parameters in the DM sector with the 
dark-matter configurations considered here. 
The results demonstrated that the interplay between hyperon-induced softening
and vector-mediated repulsion could be probed through a
combined analysis of the mass--radius relation, tidal deformability,
and moment of inertia. These findings suggested that future simultaneous constraints on
neutron-star masses and radii, gravitational-wave tidal
deformabilities, and measurements of the moment of inertia could 
provide complementary information about the high-density EOS and
the allowed properties of vector-portal dark matter.

\section*{ACKNOWLEDGMENTS}
HM and DK would like to thank Indian Institute of Technology, Bhilai for providing an exciting scientific environment during a short visit where this work was completed. DK would also like to express his gratitude for the warm hospitality extended to him at Kamala Nibas, Bhubaneswar where a part of the present work was initiated. 
SP acknowledges the Institute of Physics, Bhubaneswar, for hospitality during his sabbatical stay, where part of this work was carried out, and the funding support from SERB, Government of India, under the MATRICS project, Grant No.~MTR/2023/000687. 
This work was partially supported by Portuguese national funds from FCT (Fundação para a Ciência e a Tecnologia, I.P., Portugal) under project 2024.16290.PEX with DOI identifier 10.54499/2024.16290.PEX.

\appendix

\section{Energy Corrections in the MQMC Model}
\label{AppendixA}
In the modified quark-meson coupling (MQMC) model, the effective baryon mass receives several corrections arising from the internal quark dynamics. Among these, the center-of-mass correction plays an important role in removing the spurious motion associated with the independent quark description inside the baryon bag. In this appendix, we summarize the formalism used to evaluate the center-of-mass correction following the prescription of Guichon {\it et al.}~\cite{Guichon1988, Guichon1996}.

\subsection{Center-of-Mass Correction}
\label{AppendixA1}

Since the constituent quarks are treated independently in the confining potential, the calculated baryon energy contains an un-physical contribution associated with the motion of the center of mass of the system. This spurious contribution must be subtracted in order to obtain the physical baryon mass.

To first order in the difference between the fixed-center and relative quark coordinates, the center-of-mass correction can be expressed as the sum of two contributions,
\begin{equation}
e_{\rm c.m.}=e_{\rm c.m.}^{(1)}+e_{\rm c.m.}^{(2)} .
\label{eq:ecm_total}
\end{equation}

The first-order contribution is given by
\begin{equation}
e_{\rm c.m.}^{(1)}
=
\sum_{i=1}^{3}
\left[
\frac{m_{q_i}}
{\displaystyle \sum_{k=1}^{3} m_{q_k}}
\,
\frac{6}
{r_{0q_i}^{\,2}
\left(
3\epsilon'_{q_i}+m'_{q_i}
\right)}
\right],
\label{eq:ecm1}
\end{equation}
where the indices $i,k=(u,d,s)$ denote the quark flavors inside the baryon. Here, $\epsilon'_{q_i}$ and $m'_{q_i}$ represent the effective quark energy and effective quark mass, respectively, while $r_{0q_i}$ corresponds to the oscillator length parameter for the quark flavor $q_i$.

The second-order contribution can be written as
\begin{widetext}
\begin{align}
e_{\rm c.m.}^{(2)}
=
\frac{a}{2}
\Bigg[
&
\frac{2}{\displaystyle \sum_k m_{q_k}}
\sum_i
m_{q_i}
\langle r_i^2\rangle
+
\frac{2}{\displaystyle \sum_k m_{q_k}}
\sum_i
m_{q_i}
\langle \gamma^0(i)\, r_i^2\rangle
\nonumber
\\[2mm]
&
-
\frac{3}
{\left(\displaystyle \sum_k m_{q_k}\right)^2}
\sum_i
m_{q_i}^2
\langle r_i^2\rangle
-
\frac{1}
{\left(\displaystyle \sum_k m_{q_k}\right)^2}
\sum_i
\langle \gamma^0(i)\, m_{q_i}^2 r_i^2\rangle
\Bigg].
\label{eq:ecm2}
\end{align}
\end{widetext}

The various expectation values appearing above are defined as
\begin{equation}
\langle r_i^2\rangle
=
\frac{
\left(
11\epsilon'_{q_i}+m'_{q_i}
\right)
r_{0q_i}^{\,2}
}
{
2\left(
3\epsilon'_{q_i}+m'_{q_i}
\right)
},
\label{eq:r2}
\end{equation}

\begin{equation}
\langle \gamma^0(i)\, r_i^2\rangle
=
\frac{
\left(
\epsilon'_{q_i}+11m'_{q_i}
\right)
r_{0q_i}^{\,2}
}
{
2\left(
3\epsilon'_{q_i}+m'_{q_i}
\right)
},
\label{eq:gamma0r2}
\end{equation}
and
\begin{equation}
\langle \gamma^0(i)\, r_j^2\rangle_{i\neq j}
=
\frac{
\left(
\epsilon'_{q_i}+3m'_{q_i}
\right)
\langle r_j^2\rangle
}
{
3\epsilon'_{q_i}+m'_{q_i}
}.
\label{eq:gamma0cross}
\end{equation}

The total center-of-mass correction obtained from Eq.~(\ref{eq:ecm_total}) is subsequently included in the effective baryon mass calculation used in the equation of state for hyperonic compact star matter.

\subsection{Pionic Correction}
\label{AppendixA2}

In addition to the confinement and center-of-mass corrections, the effective baryon mass receives an important contribution from the pion cloud surrounding the baryon. The pionic correction restores, at least partially, the chiral symmetry that is broken at the mean-field level in the quark confinement approach. Following the standard MQMC formalism, the pionic self-energy contribution is evaluated perturbatively.

The pseudo-vector nucleon--pion coupling constant $f_{NN\pi}$ is determined from the Goldberger--Treiman relation,
\begin{equation}
\sqrt{4\pi}\,
\frac{f_{NN\pi}}{m_\pi}
=
\frac{g_A(N)}
{2f_\pi},
\label{eq:GTrelation}
\end{equation}
where $m_\pi$ denotes the pion mass and $f_\pi$ is the pion decay constant. The axial-vector coupling constant $g_A(N)$ in the present model is obtained as
\begin{equation}
g_A(n \rightarrow p)
=
\frac{5}{9}
\,
\frac{
5\epsilon_u^{\prime}+7m_u^{\prime}
}{
3\epsilon_u^{\prime}+m_u^{\prime}
}.
\label{eq:gA}
\end{equation}

The pionic correction to the nucleon mass is then given by
\begin{equation}
\delta M_N^\pi
=
-
\frac{171}{25}
\, f_{NN\pi}^2
\, I_\pi ,
\label{eq:pionN}
\end{equation}
where the pion loop integral $I_\pi$ is defined as
\begin{equation}
I_\pi
=
\frac{1}{\pi m_\pi^2}
\int_0^\infty
dk \,
\frac{
k^4\, u^2(k)
}{
w_k^2
},
\label{eq:Ipi}
\end{equation}
with
\begin{equation}
w_k = \sqrt{k^2+m_\pi^2}.
\label{eq:wk}
\end{equation}

The axial-vector nucleon form factor entering the pion loop integral is expressed as
\begin{equation}
u(k)
=
\left[
1
-
\frac{3}{2}
\frac{k^2}
{
\lambda_q
\left(
5\epsilon_q^{\prime}
+
7m_q^{\prime}
\right)
}
\right]
\exp\left(-\frac{k^2 r_0^2}{4}\right),
\label{eq:uk}
\end{equation}
where $\lambda_q$, $\epsilon_q^\prime$, and $m_q^\prime$ are the effective quark model parameters appearing in the confined quark wave functions.

The pionic self-energy corrections for the hyperons are similarly evaluated. For the $\Sigma^0$ and $\Lambda^0$ hyperons, the corrections are given by
\begin{equation}
\delta M_{\Sigma^0}^{\pi}
=
-
\frac{12}{5}
\, f_{NN\pi}^2 I_\pi ,
\label{eq:Sigma0pi}
\end{equation}

\begin{equation}
\delta M_{\Lambda^0}^{\pi}
=
-
\frac{108}{25}
\, f_{NN\pi}^2 I_\pi .
\label{eq:Lambda0pi}
\end{equation}

Similarly, for the charged $\Sigma$ hyperons one obtains
\begin{equation}
\delta M_{\Sigma^{\pm}}^{\pi}
=
-
\frac{12}{5}
\, f_{NN\pi}^2 I_\pi .
\label{eq:Sigmapmpi}
\end{equation}

For the cascade hyperons $\Xi^0$ and $\Xi^{-}$, the pionic corrections become
\begin{equation}
\delta M_{\Xi^{0},\Xi^{-}}^{\pi}
=
-
\frac{27}{25}
\, f_{NN\pi}^2 I_\pi .
\label{eq:Xipi}
\end{equation}

These pionic corrections are incorporated into the effective baryon masses and subsequently influence the equation of state and structural properties of hyperonic compact stars.
\subsection{One-Gluon Exchange Correction}
\label{AppendixA3}

In the MQMC model, the residual interaction between confined quarks is described through one-gluon exchange (OGE). The gluonic interaction provides an important contribution to the baryon mass spectrum and generates spin-dependent hyperfine splitting among the baryons. The interaction Lagrangian density associated with the quark--gluon coupling is given by
\begin{equation}
\mathcal{L}_{I}^{g}
=
\sum_i
J_i^{\mu a}(x)\,
A_\mu^{a}(x),
\label{eq:gluonlag}
\end{equation}
where $A_\mu^{a}(x)$ represents the octet gluon field and $J_i^{\mu a}(x)$ denotes the color current of the $i$-th quark. The quark color current is defined as
\begin{equation}
J_i^{\mu a}(x)
=
g_c\,
\bar{\psi}_q(x)\,
\gamma^\mu
\lambda_i^a
\psi_q(x),
\label{eq:colorcurrent}
\end{equation}
with $\lambda^a$ being the Gell-Mann matrices of the $SU(3)$ color group and $\alpha_c=g_c^2/4\pi$ denoting the strong coupling constant.

The gluonic correction to the baryon mass can be separated into two parts corresponding to the color-electric and color-magnetic interactions,
\begin{equation}
(\Delta E_B)_g
=
(\Delta E_B)_g^{E}
+
(\Delta E_B)_g^{M}.
\label{eq:gluon_total}
\end{equation}

The color-electric contribution is expressed as
\begin{align}
(\Delta E_B)_g^{E}
=
\frac{1}{8\pi}
\sum_{i,j}
\sum_{a=1}^{8}
\int
\frac{
d^3r_i\, d^3r_j
}{
|\vec r_i-\vec r_j|
}
\,
\langle B|
J_i^{0a}(\vec r_i)\,
J_j^{0a}(\vec r_j)
|B\rangle ,
\label{eq:gluonE}
\end{align}
whereas the color-magnetic contribution takes the form
\begin{align}
(\Delta E_B)_g^{M}
=
-
\frac{1}{8\pi}
\sum_{i,j}
\sum_{a=1}^{8}
\int
\frac{
d^3r_i\, d^3r_j
}{
|\vec r_i-\vec r_j|
}
\,
\langle B|
\vec J_i^{\,a}(\vec r_i)\cdot
\vec J_j^{\,a}(\vec r_j)
|B\rangle .
\label{eq:gluonM}
\end{align}

Using the color algebra relations for baryons,
\begin{equation}
\left\langle
\sum_a
(\lambda_i^a)^2
\right\rangle
=
\frac{16}{3},
\qquad
\left\langle
\sum_a
\lambda_i^a\lambda_j^a
\right\rangle_{i\neq j}
=
-\frac{8}{3},
\label{eq:coloralg}
\end{equation}
the gluonic corrections can be simplified into compact analytical forms.

The color-electric contribution becomes
\begin{equation}
(\Delta E_B)_g^{E}
=
\alpha_c
\left(
b_{uu}I_{uu}^{E}
+
b_{us}I_{us}^{E}
+
b_{ss}I_{ss}^{E}
\right),
\label{eq:enge}
\end{equation}
while the color-magnetic contribution is given by
\begin{equation}
(\Delta E_B)_g^{M}
=
\alpha_c
\left(
a_{uu}I_{uu}^{M}
+
a_{us}I_{us}^{M}
+
a_{ss}I_{ss}^{M}
\right),
\label{eq:engm}
\end{equation}
where $a_{ij}$ and $b_{ij}$ are numerical coefficients determined by the spin-flavor structure of the baryon under consideration.

The integrals appearing above are defined as
\begin{align}
I_{ij}^{E}
&=
\frac{16}{3\sqrt{\pi}}
\frac{1}{R_{ij}}
\left[
1
-
\frac{\alpha_i+\alpha_j}{R_{ij}^2}
+
\frac{3\alpha_i\alpha_j}{R_{ij}^4}
\right],
\label{eq:Ie}
\\[2mm]
I_{ij}^{M}
&=
\frac{256}{9\sqrt{\pi}}
\frac{1}{R_{ij}^3}
\frac{1}{
(3\epsilon_i'+m_i')
}
\frac{1}{
(3\epsilon_j'+m_j')
},
\label{eq:Im}
\end{align}
with
\begin{align}
R_{ij}^2
&=
3
\left[
\frac{1}{\epsilon_i'^2-m_i'^2}
+
\frac{1}{\epsilon_j'^2-m_j'^2}
\right],
\label{eq:Rij}
\\[2mm]
\alpha_i
&=
\frac{1}{
(\epsilon_i'+m_i')
(3\epsilon_i'+m_i')
}.
\label{eq:alphai}
\end{align}

For the nucleon, the color-electric contribution vanishes,
\begin{equation}
(\Delta E_N)_g^{E}=0,
\end{equation}
and therefore the gluonic correction reduces to the magnetic part only,
\begin{equation}
(\Delta E_N)_g^{M}
=
-
\frac{256\alpha_c}{3\sqrt{\pi}}
\left[
\frac{1}{
(3\epsilon_u'+m_u')^2
R_{uu}^3
}
\right].
\label{eq:nucleon_gluon}
\end{equation}

For the charged hyperons $\Sigma^{+}$ and $\Sigma^{-}$, the color-electric contribution is
\begin{widetext}
\begin{align}
(\Delta E_{\Sigma^\pm})_g^{E}
=
\alpha_c
\frac{16}{3\sqrt{\pi}}
\Bigg[
&
\frac{1}{R_{uu}}
\left(
1-\frac{2\alpha_u}{R_{uu}^2}
+\frac{3\alpha_u^2}{R_{uu}^4}
\right)
-\frac{2}{R_{us}}
\left(
1-\frac{\alpha_u+\alpha_s}{R_{us}^2}
+\frac{3\alpha_u\alpha_s}{R_{us}^4}
\right)
+\frac{1}{R_{ss}}
\left(
1-\frac{2\alpha_s}{R_{ss}^2}
+\frac{3\alpha_s^2}{R_{ss}^4}
\right)
\Bigg],
\label{eq:sigmaE}
\end{align}
\end{widetext}
while the corresponding magnetic contribution is
\begin{align}
(\Delta E_{\Sigma^\pm})_g^{M}
=
\frac{256\alpha_c}{9\sqrt{\pi}}
\Bigg[
&
\frac{1}{
(3\epsilon_u'+m_u')^2
R_{uu}^3
}
\nonumber\\
&
-
\frac{4}{
R_{us}^3
(3\epsilon_u'+m_u')
(3\epsilon_s'+m_s')
}
\Bigg].
\label{eq:sigmaM}
\end{align}

The total gluonic correction for $\Sigma^\pm$ is therefore
\begin{equation}
(\Delta E_{\Sigma^\pm})_g
=
(\Delta E_{\Sigma^\pm})_g^{E}
+
(\Delta E_{\Sigma^\pm})_g^{M}.
\label{eq:sigmatotal}
\end{equation}

The expressions for $\Sigma^0$ are identical to those of $\Sigma^\pm$,
\begin{equation}
(\Delta E_{\Sigma^0})_g
=
(\Delta E_{\Sigma^0})_g^{E}
+
(\Delta E_{\Sigma^0})_g^{M}.
\label{eq:sigma0total}
\end{equation}

For the $\Lambda$ hyperon, the color-electric contribution remains the same as that of $\Sigma^0$,
\begin{equation}
(\Delta E_{\Lambda})_g^{E}
=
(\Delta E_{\Sigma^0})_g^{E},
\label{eq:lambdaE}
\end{equation}
whereas the magnetic contribution differs because of the distinct spin-flavor configuration,
\begin{equation}
(\Delta E_{\Lambda})_g^{M}
=
-
\frac{256\alpha_c}{3\sqrt{\pi}}
\left[
\frac{1}{
(3\epsilon_u'+m_u')^2
R_{uu}^3
}
\right].
\label{eq:lambdaM}
\end{equation}

Thus, the total gluonic correction for the $\Lambda$ hyperon becomes
\begin{equation}
(\Delta E_{\Lambda})_g
=
(\Delta E_{\Lambda})_g^{E}
+
(\Delta E_{\Lambda})_g^{M}.
\label{eq:lambdatotal}
\end{equation}

For the cascade hyperons $\Xi^0$ and $\Xi^{-}$, the color-electric contribution remains identical to that of $\Sigma^0$ and $\Lambda$, while the magnetic contribution is given by
\begin{align}
(\Delta E_{\Xi})_g^{M}
=
\frac{256\alpha_c}{9\sqrt{\pi}}
\Bigg[
&
\frac{1}{
(3\epsilon_s'+m_s')^2
R_{ss}^3
}
\nonumber\\
&
-
\frac{4}{
R_{us}^3
(3\epsilon_u'+m_u')
(3\epsilon_s'+m_s')
}
\Bigg].
\label{eq:xiM}
\end{align}

Finally, the total gluonic correction for the $\Xi$ hyperons is obtained as
\begin{equation}
(\Delta E_{\Xi})_g
=
(\Delta E_{\Xi})_g^{E}
+
(\Delta E_{\Xi})_g^{M}.
\label{eq:xitotal}
\end{equation}

These gluonic corrections are subsequently incorporated into the effective baryon masses entering the equation of state for hyperonic compact star matter.

\bibliographystyle{apsrev4-2}
\bibliography{zzbib}
\bibliographystyle{utcaps_mod}
\end{document}